\documentclass[11pt]{article}

\usepackage[final]{acl}
\usepackage{authblk}
\usepackage[T1]{fontenc}
\usepackage[utf8]{inputenc}
\usepackage{microtype}
\usepackage{inconsolata}         % 等宽字体
\usepackage{times}               % 仅保留一次，或用 newtxtext,newtxmath
\usepackage{graphicx}
\usepackage{booktabs}
\usepackage{makecell}
\usepackage{multirow}
\usepackage{array}               % 增强表格列格式
\usepackage[table]{xcolor}       % 表格颜色支持
\usepackage{colortbl}            % 与 xcolor[table] 配合

\usepackage{amsmath}
\usepackage{amssymb}             % 包含 amsfonts
\usepackage{amsthm}
\usepackage{braket}              % 量子括号

\usepackage{enumitem}
\setlist[itemize]{topsep=1pt, itemsep=1pt, parsep=0pt, partopsep=0pt}
\setlist[enumerate]{topsep=1pt, itemsep=1pt, parsep=0pt, partopsep=0pt}

\usepackage[font=small]{caption}
\usepackage{soul}
\usepackage{url}
\hypersetup{
    hidelinks
}

\usepackage{algorithm}
\usepackage{algorithmic}         % 或 algpseudocode，二选一
\usepackage[most]{tcolorbox}
\usepackage{varwidth}

\newtcolorbox{promptbox}[1]{%
  enhanced,
  breakable,
  colback=gray!3,
  colframe=gray!45,
  boxrule=0.5pt,
  arc=2pt,
  left=6pt, right=6pt, top=5pt, bottom=5pt,
  title=\textbf{#1},
  fonttitle=\normalsize,
  before skip=4pt,
  after skip=6pt
}

\usepackage{pifont}
\newcommand{\cmark}{\ding{51}}
\newcommand{\xmark}{\ding{55}}

\usepackage[switch]{lineno}
\makeatletter
\newcommand{\runinsubsubsection}[1]{%
  \par\addvspace{4pt plus 2pt minus 1pt}%
  \noindent{\normalsize\bf\raggedright #1}\@addpunct{.}\enspace%
  \ignorespaces
}
\makeatother

\title{HCP-MAD: Heterogeneous Consensus-Progressive Reasoning for \\ Efficient Multi-Agent Debate}

\author[1]{Yiqing Liu}
\author[1]{Hantao Yao}
\author[1]{Wu Liu}
\author[ ]{Allen He} % 关键：加上空的方括号
\author[1]{Yongdong Zhang}

\affil[1]{University of Science and Technology of China}

\begin{document}
\maketitle
\begin{abstract}
Multi-Agent Debate (MAD) is a collaborative framework in which multiple agents iteratively refine solutions through the generation of reasoning and alternating critique cycles.
% Current work primarily optimizes intra-round topologies and inter-round interactions
% separately, which still results in high token costs regardless of task complexity. 
Current work primarily optimizes intra-round topologies and inter-round interactions separately, limiting the adaptation of token costs to task complexity.
This work introduces Heterogeneous Consensus-Progressive Reasoning for Efficient Multi-Agent Debate (HCP-MAD), leveraging consensus as a dynamic signal to facilitate progressive reasoning. 
The core motivation is that a majority of straightforward tasks can be effectively resolved via lightweight pair-agent debates, while complex tasks require expanded collaboration.  
% Consequently, HCP-MAD employs a three-stage progressive reasoning mechanism to develop adaptive solutions across varying task complexities. 
Firstly, Heterogeneous Consensus Verification conducts rapid consensus verification using a pair of heterogeneous agents for early stopping.
Next, Heterogeneous Pair-Agent Debate applies an adaptive stopping criterion to terminate mutual 
critique of reasoning traces.
Finally, the unresolved tasks are addressed through Escalated Collective Voting by aggregating diverse perspectives from additional agents.
Experiments across six benchmarks show that HCP-MAD enhances accuracy while substantially reducing token costs.\footnote{Code is \url{https://github.com/fuyu66/HCP-MAD}.}
\end{abstract}

\section{Introduction}

With the development of Large Language Models (LLMs), Single-Agent systems (SAS)  have demonstrated remarkable capabilities across diverse tasks ~\cite{kojima2022large,wan2023better,li2025fundamental}. 
Based on the general knowledge contained in LLMs, SAS employs a unified and straightforward reasoning process to generate responses to each task, albeit with limited capability for handling dynamic or highly complex tasks~\cite{mirzadeh2024gsm,jiang2024peek,guo2024large}.
Chain-of-thought (CoT) prompting~\cite{wei2022chain}, self-consistency~\cite{wangself,liescape}, and self-reflection~\cite{madaan2023self} further enhance SAS by guiding reasoning steps, aggregating samples, or enabling iterative refinement.
% To address these limitations, some intra-agent enhancement strategies are proposed to guide intermediate steps, aggregate samples, or enable iterative refinement, \emph{e.g.,} chain-of-thought (CoT) prompting~\cite{wei2022chain}, self-consistency~\cite{wangself,liescape}, and self-reflection~\cite{madaan2023self}.
Nevertheless, these approaches remain constrained by the limited diversity of reasoning, leading to poor performance.

\begin{figure}
	\centering
	\includegraphics[width=0.95\linewidth]{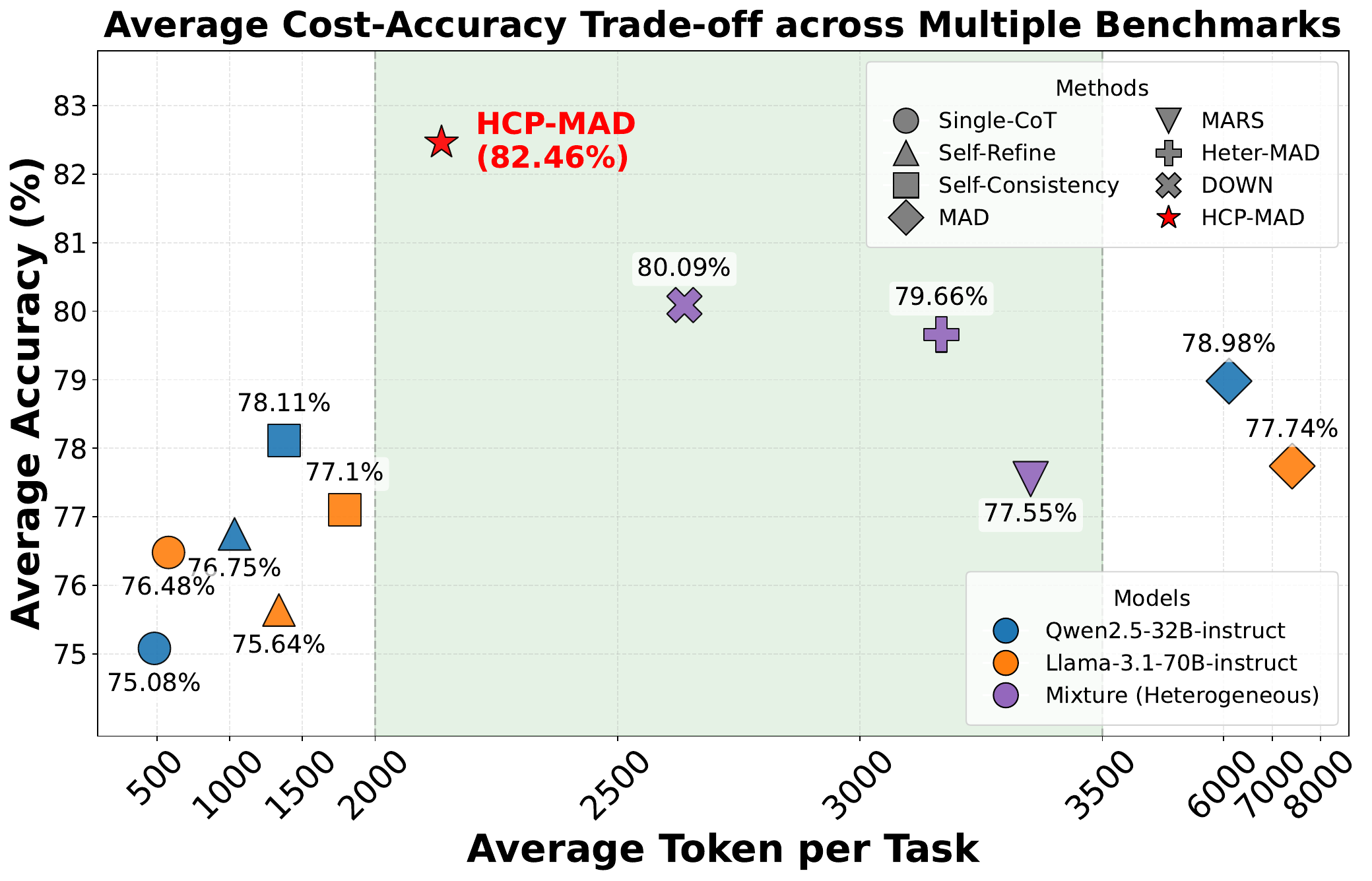}
    \caption{Comparison of HCP-MAD and existing methods regarding performance and cost averaged on six benchmarks. }
 \label{fig:1}
%\vspace{-1.5em}
\end{figure}

Recently, Multi-Agent Systems (MAS) have been introduced by leveraging information exchange, mutual critique, and iterative refinement~\cite{li2023theory,chen2023agentverse,wu2024autogen}.
A straightforward motivation is a multi-agent voting mechanism that uses the majority voting~\cite{wangself,jiang2023llm,liang2024encouraging} or weighted averaging strategies~\cite{li2022advance,du2023improving} to aggregate the responses generated by multiple agents independently.
However, it lacks the interaction among different agents, failing to resolve shared biases or deeper cognitive conflicts. 
Multi-Agent Debate (MAD) involves iteratively critiquing and refining intermediate solutions to facilitate the exchange of thoughts among agents~\cite{chan2023chateval,liang2024encouraging,fang2025counterfactual}. 
However, many MAD methods~\cite{wang2024rethinking,snell2025scaling} employ a debate process with fixed interaction topologies and a predetermined number of rounds for all tasks, resulting in token redundancy and inaccuracies due to overfitting the debate.
To enhance the efficiency of MAD, some recent studies~\cite{li2024improving,li2025fundamental} aim to generate optimized intra-round topologies by refining the communication structure. 
Moreover, others~\cite{eo2025debate,fan2025imad} focus on inter-round dynamics, employing self-adaptive mechanisms to learn optimal round counts or termination conditions.
However, interaction topologies and rounds dynamics should not be optimized independently, since the former controls the breadth of information exchange and the latter controls the depth of iterative refinement.
The key challenge is to balance these two dimensions across tasks, avoiding both costly over-collaboration on simple cases and ineffective prolonged debate on complex ones.
% However, these two dimensions are coupled: the topology determines how much new information each round can introduce, while the round-control mechanism determines whether the current topology is still worth using. Therefore, it is crucial to coordinate both aspects within a unified framework for an efficient MAD system.

% To address this challenge, an efficient MAD framework should avoid unnecessary debate for simple tasks, continue lightweight interaction only when it remains informative, and expand collaboration for complex tasks that require additional evidence.
% Despite these advancements, existing MAD methods remain inefficient,
% as they typically optimize interaction topologies or debate rounds individually.
% Therefore, it is crucial to jointly optimize both aspects within a unified framework for proposing an effective MAD system.

\begin{figure}
	\centering
    \includegraphics[width=0.9\linewidth, height=4.2cm]{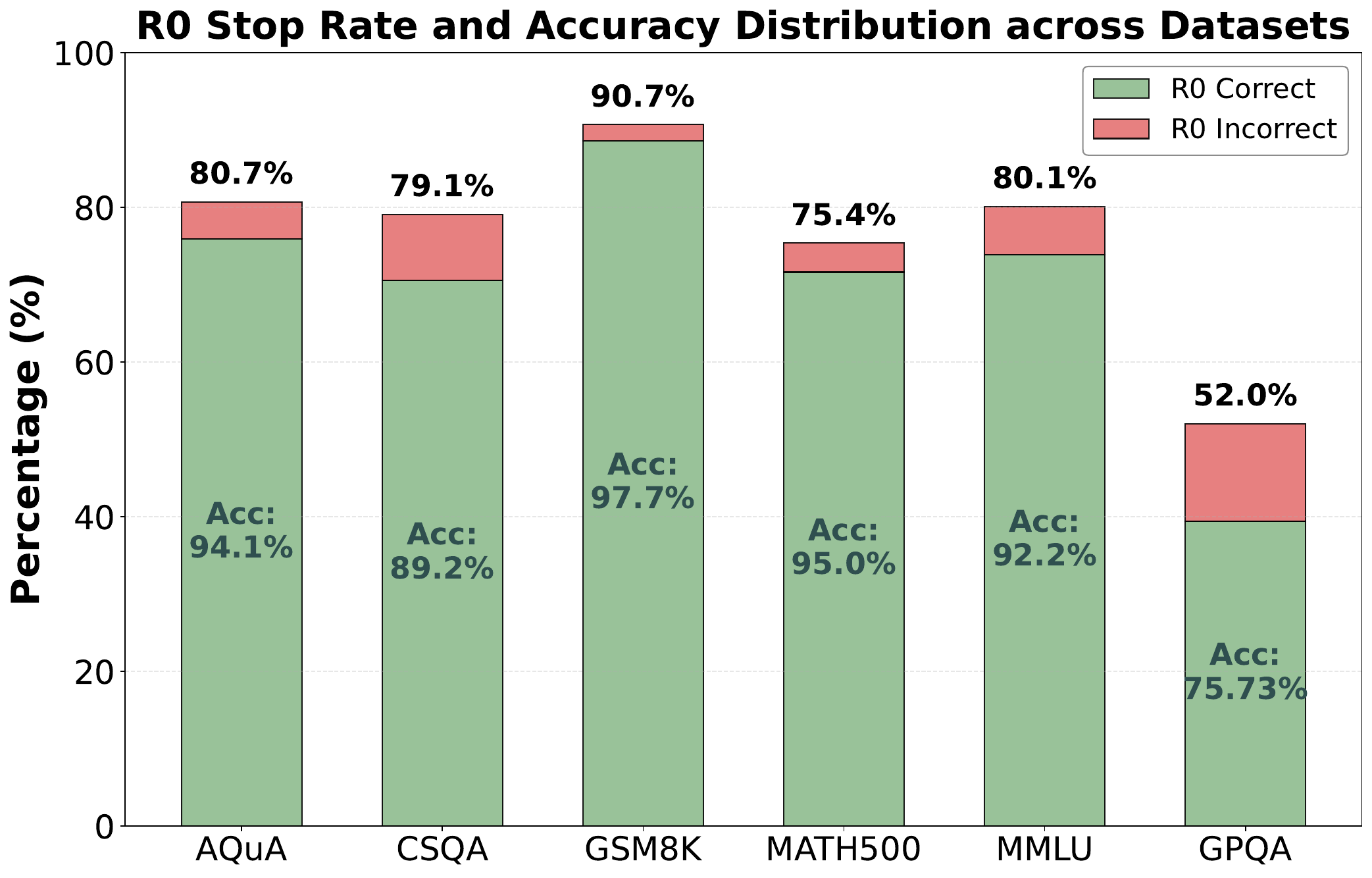}
\caption{Distribution of R0 stop rates and accuracy. High stop rates (bar height) indicate effective task coverage, while the green sections validate the effectiveness of consensus-based early stopping.}
 \label{fig:3}
  \vspace{-1.2em}
\end{figure}

% Guided by this perspective, an efficient MAD framework should allocate collaboration progressively according to task difficulty.
% It should start from a lightweight topology for straightforward tasks, continue debate only when refinement remains useful, and introduce broader aggregation when further debate provides limited information gain.
% To address the above limitations, an efficient MAD framework should avoid unnecessary debate for simple tasks, while solving the complex tasks with expanded collaboration.  
To address the above limitations, an efficient MAD framework should avoid unnecessary debate for simple tasks, while solving the complex tasks with expanded collaboration.  
Therefore, we introduce a unified and effective debate framework grounded in two key insights.
Firstly, \emph{the heterogeneous pair-agent interaction provides a lightweight topology for efficient consensus verification.}
%Firstly, \emph{the heterogeneous pair-agent debates exhibit efficiency in achieving a consensus for a majority of tasks.}
As the most lightweight multi-agent interactive topology, a pair of heterogeneous agents effectively resolve the majority of reasoning tasks. 
This efficiency stems from complementary inductive biases, which break the echo-chamber effect common in homogeneous groups and make the resulting consensus more reliable as a stopping signal. 
Crucially, when such a pair-agent reaches consensus, which may typically occur within one round, the accuracy of the consensual answer exceeds 75\%, confirming that rapid pairwise consensus is both a reliable and computationally efficient indicator for terminating reasoning on simpler tasks, as shown in Figure~\ref{fig:3}.
%Secondly, \emph{an escalated voting mechanism is superior to the debate mechanism for some complex tasks.} 
Secondly, \emph{unresolved debates often require broader evidence rather than simply more debate.}
Although larger debate topologies can introduce additional agents, Figure~\ref{fig:4} shows that increasing the number of debating agents yields limited accuracy gains but substantially higher token consumption.
Moreover, consistent with observations in prior MAD studies~\cite{wang2024rethinking,zhang2025stop}, the continued debate within a fixed group for complex tasks is not consistently effective.
For many complex tasks, agents tend to fall into answer exchange or persistent deadlock, where the performance benefit typically saturates and may even reverse after a few rounds.
Consequently, escalating voting to aggregate more agent judgments cuts through circular debates and produces more reliable solutions.

% Consequently, escalating voting to aggregate independent judgments from more agents cuts through circular debates and produces more reliable solutions.

Motivated by these insights, we propose a novel Heterogeneous Consensus-Progressive Reasoning for Efficient Multi-Agent Debate (HCP-MAD) system, which harnesses heterogeneous consensus enabling adaptively progressive reasoning.
%As illustrated in Figure~\ref{fig:2}, 
The Heterogeneous Consensus Verification(HCV) component generates responses from two independent heterogeneous agents, subsequently evaluating the consistency of their answers. 
Upon reaching consensus, the system is designed to terminate early, thereby avoiding unnecessary token costs. 
If consensus is not achieved, HCP-MAD retains the initial reasoning traces and transitions to the debate stage.
We propose the Heterogeneous Pair-Agent Debate (HPAD) to allow agents to critique each other's reasoning. 
The debate is iteratively guided by an adaptive stopping criterion, including reaching consensus, hitting a predefined maximum number of rounds, or detecting abnormal answer generation. 
If HPAD fails to produce a consensus, HCP-MAD escalates to Escalated Collective Voting (ECT), enlisting many additional agents to aggregate their independent judgments through weighted voting for generating a final answer.

As shown in Figure~\ref{fig:1}, extensive evaluation on six benchmarks proves the effectiveness of the proposed HCP-MAD in terms of accuracy and token consumption
Our contributions can be summarized as: 1) We argue that the Heterogeneous Pair-Agent Debate mechanism can efficiently achieve consensus using a lightweight pair of heterogeneous agents for reducing token consumption. 2) We propose an efficient Heterogeneous Consensus-Driven Progressive Multi-Agent Debate (HCP-MAD) for leveraging consensus consistency as a dynamic signal to facilitate progressive reasoning.

\section{Related Work}

\paragraph{Single-Agent Systems.}
Large Language Models (LLMs) have shown strong reasoning abilities within Single-Agent Systems (SAS), where a single model independently completes every step of the reasoning process~\cite{kojima2022large,wan2023better,mirzadeh2024gsm}.
%However, this direct reasoning approach often lacks diversity, which limits its ability to handle dynamic or complex tasks.
However, this direct approach often lacks diversity and struggles with complex or dynamic tasks.
To address this, intra-agent strategies like Chain-of-Thought (CoT)~\cite{wei2022chain} and Self-Consistency~\cite{wangself} have been introduced to explicitly structure reasoning and aggregate diverse paths. Furthermore, advanced frameworks such as Self-Reflection~\cite{madaan2023self}, Tree-of-Thoughts (ToT)~\cite{yao2023tree}, and Graph-of-Thoughts (GoT)~\cite{besta2024graph} extend these capabilities by enabling iterative refinement, structured multi-path exploration, and more flexible graph-based reasoning aggregation.
%Despite these advancements, SAS still face the challenge of limited reasoning diversity, which hinders their effectiveness on more complex tasks.

\paragraph{Multi-Agent Systems and Multi-Agent Debate.} 

Multi-Agent Systems (MAS)~\cite{li2023theory,wu2024autogen} extend the capabilities of LLMs by leveraging collective intelligence through collaboration and information exchange. 
A representative MAS is  Multi-Agent Debate (MAD)~\cite{chan2023chateval,liang2024encouraging} in which agents iteratively critique and refine reasoning. 
Some approaches~\cite{chan2023chateval,chen2024reconcile} employ iterative feedback and judicial roles to improve accuracy, yet their reliance on static all-to-all communication leads to computational redundancy and premature consensus.
Recently, MAD has been improved by optimizing the intra-round topology and inter-round interaction. 
For example, MARS~\cite{wang2025mars} and GroupDebate~\cite{liu2024groupdebate} restructure interaction into hierarchical or clustered workflows, and Heter-MAD~\cite{zhang2025stop} introduces model heterogeneity to maintain diversity. 
Otherwise, DOWN~\cite{eo2025debate} and iMAD~\cite{fan2025imad} employ adaptive triggers or trained discriminators to skip unnecessary rounds.
%Notably, recent studies exhibit a profound divergence regarding the necessity of consensus.
%Significantly, recent studies diverge on the necessity of consensus: 
Differently, Aegean~\cite{ruan2025reaching} formalizes refinement as a distributed problem to ensure reliable termination within a fixed agent group, and Free-MAD~\cite{cui2025free} utilizes the anti-conformity prompts and trajectory scoring to avoid majority bias.
%HCP-MAD keeps the consensus idea but turns consensus into a guiding signal that actively adjusts reasoning depth across adaptive stages, enabling precise resource control for tasks of different complexity.
In summary, most existing methods improve topology and interaction separately. 
The static topologies fail to dynamically reason, and existing dynamic stopping rules neglect the potential for structural reconfiguration to accommodate evolving reasoning needs.
%structural changes stay fixed during reasoning, and dynamic stopping rules do not fully consider the changing needs of the agent group. 
%This limits adaptation to task complexity. 
Based on the fact that the heterogeneous pair-agent debates exhibit efficiency in achieving a consensus for a majority of tasks, we propose a novel HCP-MAD with the lightweight debate structure and adaptive reasoning progression to boost reasoning while reducing token consumption.
%which jointly controls reasoning progression and depth to achieve the best balance between accuracy and efficiency.

%\emph{Vanilla Majority-Voting.}

\section{Methodology}

\subsection{System Definition}

A multi-agent debate (MAD)~\cite{li2023theory,kim2025towards} is defined as a tuple $\mathcal{S} = \langle \mathcal{A}, \mathcal{Q}, \mathcal{M} \rangle$, where $\mathcal{A} = \{a_1, \dots, a_N\}$ is a set of $N$ agents, $\mathcal{Q}$ represents the set of query tasks, and $\mathcal{M}$ represents the Large Language Model (LLM) set used to initialize the agents.
Given a query $q \in \mathcal{Q}$, the agent $a_i$ generates a response $\boldsymbol{r}_{i,t}$ at $t$-th round,
\begin{equation}
    \boldsymbol{r}_{i,t} = \Psi_i \bigl( q, P_i, \mathcal{H}_t \bigr), %t\in[1,T],
\end{equation}
where $\Psi_i \in \mathcal{M}$ is instantiated by the LLM,  $P_i$ is the system prompt,
and \(\mathcal{H}_t\) is the observable history available before step \(t\).
After that, a task-specific function $\mathcal{J}(\cdot)$ extracts the predicted answer $\hat{y}_{i,t}$ from the response $\boldsymbol{r}_{i,t}$,
\begin{equation}
\hat{y}_{i,t} = \mathcal{J}(\boldsymbol{r}_{i,t}).
\end{equation}

Based on the answer from all agents, let $\hat{y}_{i,T}$ denote the answer of the agent $a_i$ at $T$-th round, a decision-making function $\mathcal{F}$ is employed to generate the final answer $\hat{y}$, 
\begin{equation}
    \hat{y} = \mathcal{F}(\{\hat{y}_{i,T}\}_{i=1}^N).
\end{equation}

To boost the efficiency of MAD, we introduce a novel \emph{Heterogeneous Consensus-Progressive Reasoning for Efficient Multi-Agent Debate (HCP-MAD)}, which can be viewed as a finite history-dependent transition system,
\begin{equation}
\mathfrak{T}_{\mathrm{HCP-MAD}}=(\mathbb{X},\mathbb{H},\delta,\Gamma,T),
\end{equation}
where $\mathbb{X}=\{\mathrm{C},\mathrm{D},\mathrm{V},\mathrm{E}\}$
is the stage space corresponding to Heterogeneous Consensus
Verification (HCV), Heterogeneous Pair-Agent Debate (HPAD),
Escalated Collective Voting (ECV), and the terminal state (END), respectively.
$\mathbb{H}$ is the space of observable reasoning histories,
%$\delta:\mathcal{X}\times\mathcal{H}\rightarrow\mathcal{X}$ 
$\delta:\mathbb{X}\times\mathbb{H}\rightarrow\mathbb{X}$ is the
stage-transition function, $\Gamma$ is the history-update operator, and
$T$ is the finite reasoning horizon. 

 \begin{figure}
	\centering
	\includegraphics[width=1.0\linewidth]{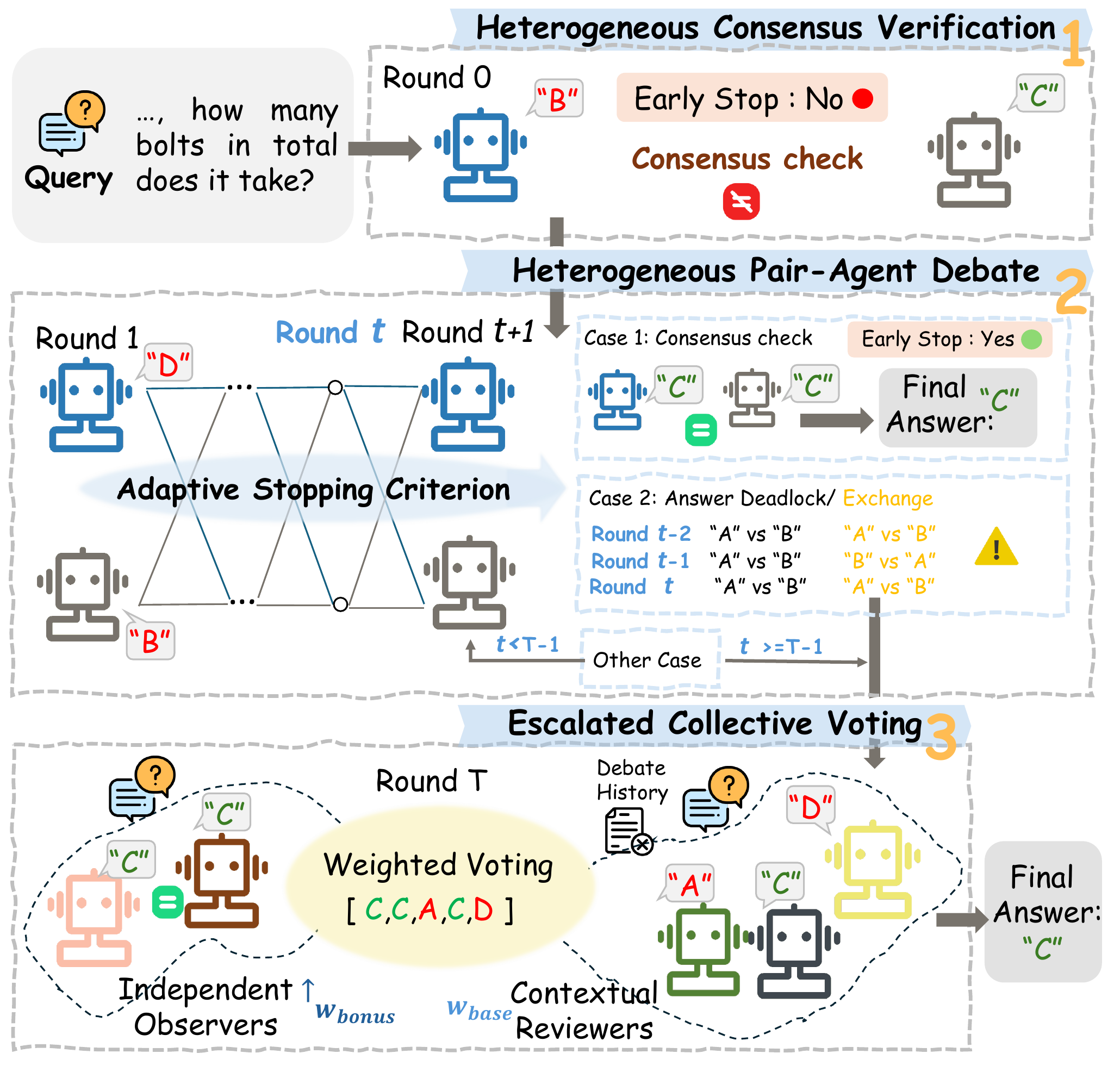}
 \caption{HCP-MAD conducts consensus-guided progressive reasoning in three stages: (1) Heterogeneous Consensus Verification (HCV) for early stopping, (2) Heterogeneous Pair-Agent Debate (HPAD) uses a lightwight pair-debate with adaptive stopping criteria to resolve conflicts; and (3) Escalated Collective Voting (ECV) employing independent observers and contextual reviewers for weighted voting.}
\label{fig:2}
\vspace{-1.0em}
\end{figure}

The system starts from $x_0=\mathrm{C}, \mathcal{H}_0=\emptyset$ .
%At step \(t\), the current history \(\mathcal{H}_t\) is used as the input
% context for generating \(\boldsymbol{r}_{i,t}\). After the current stage
% \(x_t\) is executed, 
At step $t$, the current stage $x_t$ invokes a subset of agents and produces a 
set of new responses, denoted as $\mathcal{R}_t(x_t,q,\mathcal{H}_t)$. 
its generated responses are
recorded into the next history \(\mathcal{H}_{t+1}\). The stage and history evolve as
\begin{equation}
\begin{aligned}
%\mathcal{H}_{t+1} &= \Gamma(\mathcal{H}_t,x_t),\\
  \mathcal{H}_{t+1} &= \Gamma\bigl(\mathcal{R}_t(x_{t},q,\mathcal{H}_t)\bigr),\\
x_{t+1} &= \delta(x_t,\mathcal{H}_{t+1}),
\end{aligned}
\qquad t=0,\dots,{T}.
\end{equation}
The admissible transitions are
\begin{equation}
C \rightarrow\{{D},{E}\},
{D}\rightarrow\{{D},{V},{E}\},
{V}\rightarrow{E}.
\end{equation}

By the above transition rules, the accuracy-cost trade-off can be summarized as, %we define a generic accuracy-cost utility:
\begin{equation}
\mathcal{U}(Q)
=
\mathbb{E}_{q\sim\mathcal{Q}}
\left[
%\text{Accuracy}(Q) - \lambda \cdot \text{AvgCost}(Q)$
\mathrm{Accuracy}(q)- \lambda\,\mathrm{Cost}(q)
\right],
\end{equation}
where %\(y^{\star}\) is the ground-truth answer, 
\(\mathrm{Cost}(q)\)
is the total token cost, 
and \(\lambda\) controls the
trade-off between answer quality and computational cost. This objective is
not used for training. It only formalizes the evaluation criterion of
HCP-MAD. 

This formulation highlights the main design principle of HCP-MAD:
many straightforward tasks can be effectively solved through lightweight pair-agent debates, while the complex tasks require expanded collaboration. Specifically, HCP-MAD leverages a consensus-guided three-stage progressive reasoning to dynamically adjust the collaboration according to task complexities, as shown in Figure \ref{fig:2}. key components are described below.

\subsection{Heterogeneous Consensus Verification}
\label{sec:stage1}

\begin{tcolorbox}[
    colframe=green!40!black,
    colback=white,
    boxrule=0.7pt,
    arc=2pt,
    boxsep=0pt,
    left=5pt,
    right=5pt,
    top=2pt,
    bottom=2pt,
    before skip=3pt,
    after skip=6pt
]
\noindent\textbf{\textcolor{green!40!black}{Finding 1:}}
The consensus between two heterogeneous agents can provide an efficient indicator of answer reliability.
\end{tcolorbox}
HCV aims to resolve straightforward tasks  cost-efficiently by identifying consistent solutions early.  Here, heterogeneity means that agents differ in backbones or other reasoning configurations, which can provide more robust epistemic diversity and complementary reasoning behaviors.
As shown in Figure~\ref{fig:3}, these agents reach consensus on many tasks with a high proportion of correct solutions, indicating that the heterogeneous consensus is a powerful indicator of answer reliability that generalizes across varied task complexities. 

HCV first invokes a minimal set of heterogeneous agents $\mathcal{A}_{init} = \{a_1, a_2\} \subset \mathcal{A}$. 
Given the query $q$ and the prompt $P_i$, the agent $a_i \in \mathcal{A}_{init}$ independently generates its initial response,
\begin{equation}
    \boldsymbol{r}_{i,0} = \Psi_i(q, P_i, \mathcal{H}_0), \quad i \in \{1, 2\},
\end{equation}
where \(\mathcal{H}_0=\emptyset\) indicates that no historical reasoning is provided.
Next, the predicted result $\hat{y}_{i,0}= \mathcal{J}(\boldsymbol{r}_{i,0})$ is extracted by the function $\mathcal{J}(\cdot)$.
The initial consensus indicator is defined as,
\begin{equation}
\Phi_{init} = \mathbb{I}\bigl( \hat{y}_{1,0} = \hat{y}_{2,0} \bigr),
\end{equation}
where $\mathbb{I}(\cdot)$ denotes the indicator function. %\hat{y_{1,0}}
HCV corresponds to the initial state $x_0=\mathrm{C}$. After HCV is executed, the history is updated from
\(\mathcal{H}_0\) to \(\mathcal{H}_1\):
\begin{equation}
\mathcal{H}_{1}
=
\Gamma\bigl(\mathcal{R}_0(x_0, q,\mathcal{H}_0)\bigr)
%\Gamma(\mathcal{H}_{0},\mathrm{C})
=
% \mathcal{H}_{0}
% \cup
\{
\boldsymbol{r}_{1,0},
\boldsymbol{r}_{2,0}
\}.
\end{equation}
The HCV transition rule is defined as,
\begin{equation}
\delta(C,\mathcal{H}_1)=
\begin{cases}
{E}, & \Phi_{init}=1,\\
D, & \Phi_{init}=0,
\end{cases}
\end{equation}
If $\Phi_{init}=1$, HCP-MAD stops early and returns the
agreed answer $\hat{y}=\hat{y}_{1,0}=\hat{y}_{2,0}$. Otherwise,$\mathcal{H}_1$ is retained as the first
non-empty history and the system transitions to HPAD.

\subsection{Heterogeneous Pair-Agent Debate}
\begin{figure}
	\centering   
	\includegraphics[width=1\linewidth]{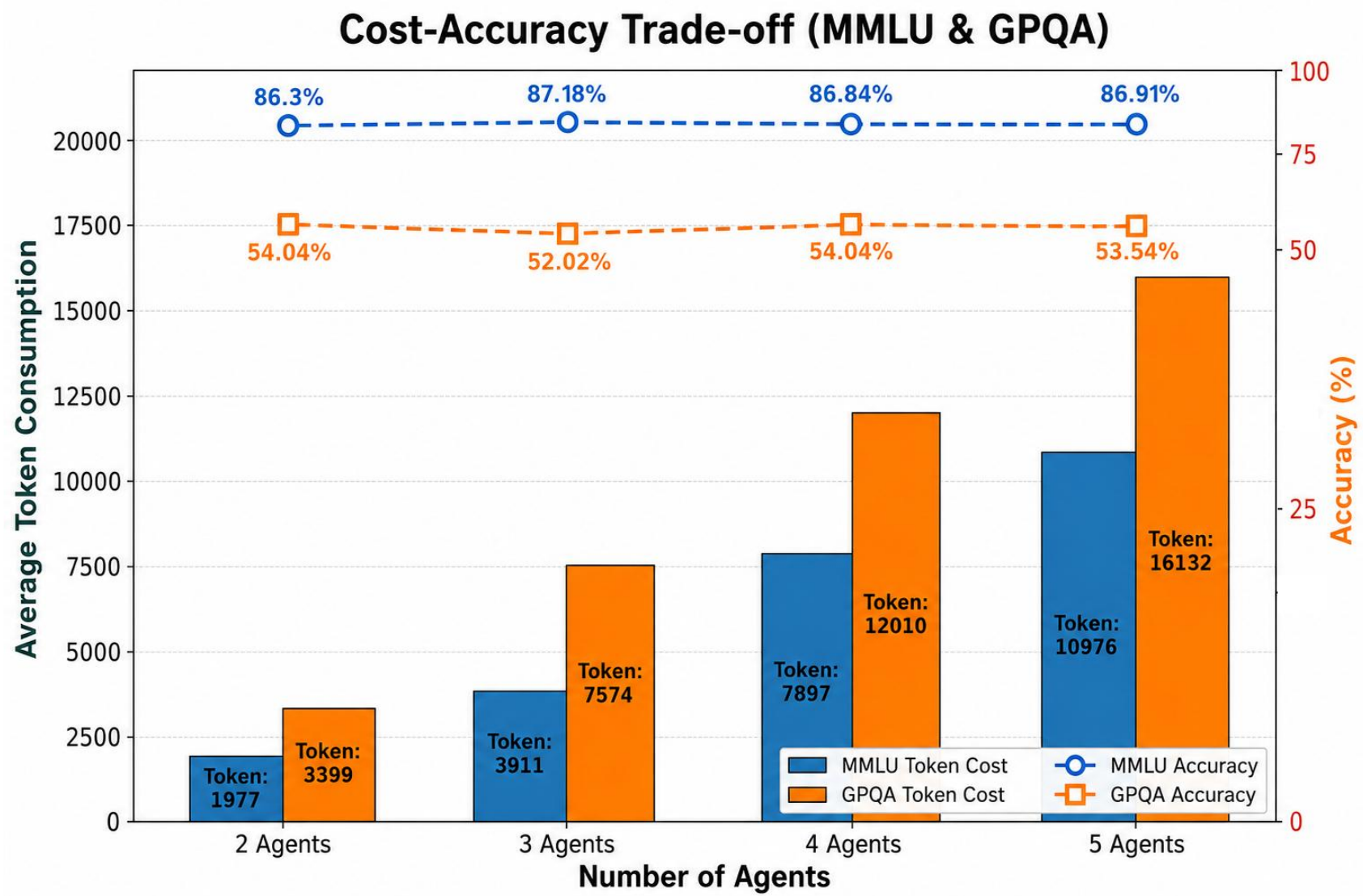}
    \caption{{Analysis of accuracy and token costs of HCP-MAD with varying agents in the debate stage.}}
 \label{fig:4}
 \vspace{-1.0em}
\end{figure}

\begin{tcolorbox}[
    colframe=green!40!black,
    colback=white,
    boxrule=0.7pt,
    arc=2pt,
    boxsep=0pt,
    left=5pt,
    right=5pt,
    top=2pt,
    bottom=2pt,
    before skip=3pt,
    after skip=6pt
]
\noindent\textbf{\textcolor{green!40!black}{Finding 2:}}
% Enlarging the debate topologies limited marginal gains under substantially higher token costs.
Enlarging the debate topology limits marginal gains while incurring substantially higher token costs.
% A heterogeneous pair is the minimal debate topology that preserves reasoning diversity while avoiding unnecessary communication overhead.
\end{tcolorbox}
While multi-agent debates can be used to solve the unresolved tasks, the token costs increase sharply with the addition of more agents and debate rounds. 
Therefore, we propose a Heterogeneous Pair-Agent Debate (HPAD) by using a lightweight debate framework with an adaptive stopping criterion to refine solutions. 
As shown in Figure~\ref{fig:4}, increasing the number of agents beyond two yields comparable performance but significantly higher token costs in HCP-MAD.

Formally, HPAD contains the heterogeneous pair agents of $\mathcal{A}_{d} = \{a_1, a_2\} \subset \mathcal{A}$, while the debate process over rounds \(t=1,\ldots,T-1\). 
At $t$-th round, given the query $q$ and the prompt $P_i$, the agent $a_i \in \mathcal{A}_{d}$  initialized with LLM $\Psi_i$ generates its response $\boldsymbol{r}_{i,t}$,
\begin{equation}
    \boldsymbol{r}_{i,t} = \Psi_i(q, P_i, \mathcal{H}_{t}), %\quad t\ge 1,
    \quad i \in \{1, 2\},
\end{equation}
where \(\mathcal{H}_t\) contains the history before round \(t\), and \(\mathcal{H}_1\) contains the initial
responses from HCV.
The predicted result is $\hat{y}_{i,t}= \mathcal{J}(\boldsymbol{r}_{i,t})$ of each agent.

To avoid unproductive debate, HPAD monitors whether debate continues to provide useful information. 
We consider two low-information patterns: \emph{Answer Exchange} and \emph{Persistent Deadlock}.
\emph{1) Answer exchange.}
If two agents repeatedly exchange answers with each other across consecutive rounds, \emph{e.g.,} $AB\!\rightarrow\!BA\!\rightarrow\!AB$, more debate rounds are unnecessary. The exchange indicator  $e_t$ as
\begin{equation}
e_t
=
\mathbb{I}\!\left(\hat{y}_{1,t}=\hat{y}_{2,t-1}\right)
\cdot
\mathbb{I}\!\left(\hat{y}_{2,t}=\hat{y}_{1,t-1}\right). %\quad t\ge 1.
\label{eq:swap-indicator}
\end{equation}
To identify persistent instability, a consecutive exchange counter $\mathcal{E}_t$ increments when an exchange occurs and resets to zero if the pattern is broken,
\begin{equation}
\mathcal{E}_t =  (\mathcal{E}_{t-1} + 1)\cdot e_t .
\label{eq:consecutive-exchange}
\end{equation}

\emph{2) Persistent Deadlock.}
If both agents keep their own answers unchanged while still disagreeing, \emph{e.g.,} $AB\!\rightarrow\!AB\!\rightarrow\!AB$, the debate has reached a deadlock. 
The current deadlock counter ${d}_t$ as,
\begin{equation}
{d}_t
=
\mathbb{I}\!\left(\hat{y}_{1,t}=\hat{y}_{1,t-1}\right)
\cdot
\mathbb{I}\!\left(\hat{y}_{2,t}=\hat{y}_{2,t-1}\right).
\label{eq:swap-indicator1}
\end{equation}
The deadlock counter $\mathcal{D}_t$ that increments only when a deadlock occurs and resets to zero immediately if the pattern is broken,

\begin{equation}
\mathcal{D}_t =(\mathcal{D}_{t-1} + 1)\cdot{d}_t .
\label{eq:consecutive-d}
\end{equation}
After round \(t\) is executed, the new responses are appended to the history:
\begin{equation}
\mathcal{H}_{t+1}
=
\Gamma\bigl(\mathcal{R}_t(x_t,q,\mathcal{H}_t)\bigr)
%\Gamma(\mathcal{H}_{t},\mathrm{D})
=
% \mathcal{H}_{t}
% \cup
\{
\boldsymbol{r}_{1,t},
\boldsymbol{r}_{2,t}
\}.
\end{equation}

The HPAD transition rule is
\begin{equation}
\delta({D},\mathcal{H}_{t+1})=
\begin{cases}
{E}, &
\hat{y}_{1,t}=\hat{y}_{2,t},\\
{V}, &
\mathcal{E}_t\geq\eta_e \ \mathrm{or}\ \mathcal{D}_t\geq\eta_d \ \mathrm{or}\ t=T-1,\\
{D}, &
\mathrm{otherwise},
\end{cases}
\end{equation}
where $\eta_e$ and $\eta_d$ are the exchange and deadlock thresholds,
respectively. In our implementation, both are set to $2$. Therefore,
HPAD continues only when pairwise debate remains informative; otherwise,
the system either terminates on consensus or escalates to ECV.

\begin{table*}[t]
\centering
\fontsize{8.5pt}{9pt}\selectfont
\renewcommand{\arraystretch}{0.9}
\setlength{\tabcolsep}{3pt}
\caption{Comparison with existing methods. Best scores are highlighted in \textbf{bold}, and second-best are \underline{underlined}. Scores for all benchmarks and Avg. Acc. are reported as accuracy percentages (\%). Avg. Tokens denotes the average number of tokens consumed per query.}
\begin{tabular}{l|cccccc|cc}
\toprule
Methods & MMLU & CommonQA & GPQA & MATH500 & GSM8K &  AQuA & \textbf{Avg. Acc.(\%)} & \textbf{Avg. Tokens} \\
\midrule
\multicolumn{9}{>{\columncolor{gray!15}}c}{Llama-3.1-70b-instruct} \\

\midrule
CoT~\cite{wei2022chain} & 76.80 & 75.27 & 48.99 & 82.20 & 90.98 & 84.65 & 76.48  & 580 \\
SR~\cite{madaan2023self} & 77.20 & 76.00 & 47.98 & 78.80 & 91.58 & 82.28 & 75.64 & 1,339 \\
SC~\cite{wangself} & 79.10 & 76.49 & 46.97 & \underline{82.40} & 92.19 & {85.43} & 77.10 & 1,792 \\
\midrule
MAD~\cite{du2023improving} & 77.95 & 78.30 & 49.49 & 82.20 & 91.89 & \underline{86.61} & 77.74 & 7,413 \\
\midrule
\multicolumn{9}{>{\columncolor{gray!15}}c}{Qwen2.5-32b-instruct} \\
\midrule
CoT~\cite{wei2022chain} & 77.61 & 84.44 & 42.93 & 78.40 & 88.17 & 80.31 & 75.08 & 483 \\
SR~\cite{madaan2023self} & 82.43 & 84.60 & 41.41 & 77.80 & 91.96 & 82.68 & 76.75 & 1,034 \\
SC~\cite{wangself} & 82.56 & \textbf{85.18} & 41.92 & 79.80 & 94.16 & 85.04 & 78.11 & 1,376 \\
\midrule
MAD~\cite{du2023improving} & 83.85 & 84.36 & 45.96 & 80.40 & 95.53 & 85.83 & 78.98 & 6,111 \\
\midrule
\multicolumn{9}{>{\columncolor{gray!15}}c}{Qwen2.5-32b-instruct + Llama-3.1-70b-instruct} \\
\midrule
MARS~\cite{wang2025mars} & 78.63 & 79.85 & 49.49 & 81.20 & 93.03 & 83.07 & 77.55 & 3,352 \\
Heter-MAD~\cite{zhang2025stop} & \underline{84.30} & 83.52 & 50.50 & 81.00 & 94.77 & 83.86 & 79.66 & 3,168 \\
DOWN~\cite{eo2025debate} & 84.06 & 84.11 & \underline{51.01} & 80.80 & \underline{95.67} & 84.65 & \underline{80.09} & 2,638 \\
\textbf{HCP-MAD(Ours)} & \textbf{86.30} & \underline{84.85} & \textbf{54.04} & \textbf{85.20} & \textbf{95.83} & \textbf{88.58} & \textbf{82.46} & 2,137 \\
\bottomrule
\end{tabular}

\label{tab:performance_comparison}
%\end{table}
%\vspace{-1.2em}
\end{table*}

\subsection{Escalated Collective Voting}
\label{sec:stage3}

For tasks that have not yet reached a consensus, 
ECV recruits additional agents to introduce broader collective intelligence and aggregate their judgments. 
Specifically, ECV enhances the diversity of reasoning outcomes by providing different heterogeneous agents with diverse reference chains of thought.
To reduce potential bias or error propagation from the debate history, we partition the escalated agents into two complementary subgroups: an \emph{Independent Subgroup  $\mathcal{A}_{\text{ind}}$} that reasons from scratch, and a \emph{Reviewer Subgroup $\mathcal{A}_{\text{rev}}$} that evaluates the debate history. 
Formally, ECV escalates a new escalation agent pool 
$\mathcal{A}_{\text{esc}}=\mathcal{A}_{\text{ind}}\cup\mathcal{A}_{\text{rev}}\subseteq\{a_3,\dots,a_N\}$, where
$\mathcal{A}_{\text{ind}}\cap\mathcal{A}_{\text{rev}}=\emptyset$.

\emph{Independent Observers ($\mathcal{A}_{\text{ind}}$)} consisting of $N_1$ additional agents provide unbiased perspectives by reasoning independently.
For the agent $a_i \in \mathcal{A}_{\text{ind}}$, 
the generated response $\boldsymbol{r}_{i,T}^{\text{(ind)}}$ and extracted candidate answer $\hat{y}_{i,T}^{\text{(ind)}}$ are defined as,
\begin{equation}
\hat{y}_{i,T}^{\text{(ind)}} = \mathcal{J}(\boldsymbol{r}_{i,T}^{\text{(ind)}})=\mathcal{J}(\Psi_i(q, P_{\text{ind}}, \emptyset)).
\end{equation}
where $P_{\text{ind}}$ is the prompt for independent reasoning.

Moreover, \emph{Contextual Reviewers ($\mathcal{A}_{\text{rev}}$)} consists of  $N_2$ agents (where $N_1 < N_2$) who serve as judges by critically analyzing the summarized debate context $\mathcal{C}_{deb}$. For the agent $a_i \in \mathcal{A}_{\text{rev}}$, the generated response $\boldsymbol{r}_{i,T}^{\text{(rev)}}$ and generated answer $\hat{y}_{i,T}^{\text{(rev)}}$ at the final round T are defined as,
\begin{equation}
\hat{y}_{i,T}^{\text{(rev)}} = \mathcal{J}(\boldsymbol{r}_{i,T}^{\text{(rev)}})=\mathcal{J}(\Psi_i(q, P_{\text{rev}}, \mathcal{C}_{deb} )),
\end{equation}
where $\mathcal{C}_{deb}=Summarize(\mathcal{H}_T)$  denotes the summary of the final debate round history, and  $P_{\text{rev}}$ is the prompt for contextual review.

The entire generated answer set can be defined as
$\{
\hat{y}_{i,T}
\}
=
\{
\hat{y}_{i,T}^{(\mathrm{ind})}
\}
\cup
\{
\hat{y}_{i,T}^{(\mathrm{rev})}
\}$.
The final system output $\hat{y}$ is a weighted majority vote over the entire candidate answer space $\mathcal{Y}$, 
\begin{equation}
    \hat{y} = \arg\max_{c \in \mathcal{Y}} \sum_{a_i \in \mathcal{A}_{\text{esc}}} w_i \cdot \mathbb{I}\bigl( \hat{y}_{i,T} = c \bigr),
\end{equation}
where $c$ is a candidate answer, and $w_i$ prioritizes the independent group only when they exhibit \emph{unanimous consensus},
\begin{equation}
    w_i = w_{\text{base}} + \underbrace{\beta \cdot \mathbb{I}\bigl( a_i \in \mathcal{A}_{\text{ind}} \land \Phi_{\text{unanimous}} = 1 \bigr)}_{\text{Independent Consensus Bonus}} ,
\label{eq:17}
\end{equation}
%where $w_{\text{base}} = 1.0$. 
where $\beta = (N_2 - N_1)/N_2$ is the bonus coefficient for adjusting the importance of independent observers only when the entire subgroup $\mathcal{A}_{\text{ind}}$ reaches an identical conclusion, and $\Phi_{\text{unanimous}}$ is the unanimous indicator defined as,
\begin{equation}
    \Phi_{\text{unanimous}} = \mathbb{I} \left( \forall a_i, a_j \in \mathcal{A}_{\text{ind}}, \hat{y}_{i,T} = \hat{y}_{j,T}\right).
\end{equation}
Finally, ECV transitions to the terminal state:
\begin{equation}
\delta(V,\cdot)={E}.
\end{equation}
% In conclusion, ECV can effectively prioritize the unanimous consensus of independent agents over potential biases in the debate history, while reverting to standard voting when disagreement persists.

\section{Experiments}
\subsection{Experimental Setup}
In the following, we give a brief description of the experimental setup, and more detailed information can be referred from the Appendix.

\noindent\textbf{Datasets.} We evaluate HCP-MAD on six benchmarks:
knowledge reasoning 
(MMLU\cite{hendrycksmeasuring}, CommonsenseQA~\cite{talmor2019commonsenseqa}, GPQA-Diamond~\cite{rein2024gpqa}), and mathematical problem-solving (MATH-500\cite{hendrycks2021measuring}, GSM8K~\cite{cobbe2021training}, AQuA~\cite{ling2017program}).

\noindent\textbf{Models.} Our core experiments utilize a heterogeneous pair of Llama-3.1-70B-Instruct and Qwen2.5-32B-Instruct. More model results can be found in the appendix.

\noindent\textbf{Baselines.} We use a variety of methods as baselines: Single-agent methods include CoT~\cite{wei2022chain}, Self-Refine (SR)~\cite{madaan2023self}, and Self-Consistency (SC)~\cite{wangself}. Multi-agent methods include standard MAD and its efficiency-optimized variants (MARS~\cite{wang2025mars}, Heter-MAD~\cite{zhang2025stop}, and DOWN~\cite{eo2025debate})..

\subsection{Main Results}
In this section, we compare the proposed HCP-MAD with several existing methods across six benchmarks, the results summarized in Table~\ref{tab:performance_comparison}. 

In the single-agent setting, CoT uses the fewest tokens but suffers from lower accuracy. 
By adding self-consistency (SC) to the Qwen model, the average accuracy improves from 75.08\% to 78.11\%, while token consumption increases from 483 to 1,376.  Multi-Agent Debate is proposed to enhance single-agent reasoning capabilities. 
As shown in Table~\ref{tab:performance_comparison}, existing MAD methods such as DOWN and Heter-MAD both achieve higher performance. For instance, DOWN obtains an accuracy of 80.09\% with lower token usage, which clearly outperforms the 78.98\% accuracy of the standard MAD. 
This shows the benefits of heterogeneous models in improving overall performance.%%%%%%%

Among all MAD methods, HCP-MAD achieves the best performance with the lowest token consumption. Compared to DOWN, HCP-MAD improves the average accuracy from 80.09\% to 82.46\% while reducing token consumption from 2,638 to 2,137. 
Moreover, HCP-MAD excels on five of the six benchmarks, further demonstrating its generality and efficiency across various tasks. 
From Table~\ref{tab:performance_comparison}, it can be observed that all existing MAD-based methods perform worse on CommonQA than some single agents like SC and SR. 
The reason is that CommonQA consists of simple commonsense questions, where debate introduces unnecessary complexity and noise instead of meaningful refinement. 
However, HCP-MAD achieves the best performance of 84.85\% among all MAD-based methods, falling just short of the 85.18\% accuracy obtained by SC.

\begin{table}[t]
\centering
\begingroup
\footnotesize
\renewcommand{\arraystretch}{1.0}
\setlength{\tabcolsep}{1.2pt}
\setlength{\aboverulesep}{0.2ex}
\setlength{\belowrulesep}{0.2ex}
\setlength{\cmidrulekern}{0.2em}
\caption{Ablation study of the critical components in HCP-MAD.}
\label{tab:ablation_study_calibrated}
%\begin{tabular}{@{}p{0.43\columnwidth}cc@{\hspace{4pt}}cc@{}}
\begin{tabular}{@{}lcc@{\hspace{5pt}}cc@{}}
\toprule
\textbf{Methods} 
& \multicolumn{2}{c}{\textbf{MMLU}} 
& \multicolumn{2}{c}{\textbf{GPQA}} \\
\cmidrule(lr){2-3} \cmidrule(lr){4-5}
& Acc. & Tok. & Acc. & Tok. \\
\midrule
\rowcolor{gray!10} 
\textbf{HCP-MAD (Full)} 
& \textbf{86.30} & \textbf{1,977} 
& \textbf{54.04} & \textbf{3,724} \\
\midrule
\multicolumn{5}{@{}l}{\textit{Model Configuration}} \\
- w/ Homo. 
& 82.77 & 2,158 & 47.98 & 3,403 \\
\midrule
\multicolumn{5}{@{}l}{\textit{Stage I: HCV}} \\
- w/o Early Stopping 
& 86.36 & 3,951 & 51.52 & 5,625 \\
\midrule
\multicolumn{5}{@{}l}{\textit{Stage II: HPAD}} \\
- w/o HPAD 
& 83.58 & 6,609 & 54.55 & 6,238 \\
- w/o Adaptive Stopping 
& 86.02 & 2,753 & 51.52 & 4,180 \\
\midrule
\multicolumn{5}{@{}l}{\textit{Stage III: ECV}} \\
- w/o ECV 
& 85.28 & 1,713 & 51.52 & 3,412 \\
- Replace Vote with Debate 
& 86.36 & 5,224 & 52.53 & 4,698 \\
- w/ Simple majority vote 
& 85.89 & 1,753 & 52.53 & 3,612 \\
- w/ Reviewer majority vote 
& 86.16 & 2,458 & 53.03 & 4,651 \\
- w/o Weighted Bonus ($\beta=0$) 
& 86.02 & 1,977 & 53.03 & 3,724 \\
\bottomrule
\end{tabular}
\endgroup
\vspace{-1.0em}
\end{table}

\subsection{Ablation Study}
In this section, we conduct extensive ablation studies to verify the effectiveness of the proposed components in HCP-MAD. 

\noindent\textbf{Effect of the Proposed Components.} 
HCP-MAD comprises three essential components: HCV, HPAD, and ECV. We analyze the contribution of each component and summarize results in Table~\ref{tab:ablation_study_calibrated}.
HCV serves as the baseline for HCP-MAD, achieving an accuracy of 82.02\% on MMLU. When HCV is combined with HPAD, the accuracy improves significantly from 82.02\% to 85.28\%. 
In contrast, HCP-MAD without HPAD achieves a performance of 83.58\%, which is lower than the 86.30\% obtained with HPAD.
This analysis underscores the importance of the pair-agent debate. 
After excluding the ECV component, HCP-MAD reaches an accuracy of 85.28\%, further emphasizing the critical role of ECV in resolving complex tasks.

\noindent\textbf{Effect of the Heterogeneous Reasoning.} 
%\noindent\textbf{Effect of the Model Heterogeneity} 
Heterogeneous reasoning aims to provide diverse perspectives to enhance the robustness of consensus. 
As shown in Table~\ref{tab:ablation_study_calibrated}, replacing heterogeneous pairs with homogeneous agents, as seen in `HCP-MAD w/ Homo.', results in a significant performance decrease from 86.30\% to 82.77\% on the MMLU. 
These results illustrate the importance of heterogeneous reasoning for reaching sound judgments through consensus.

\noindent\textbf{Effect of Heterogeneous Pair-Agent Debate.}
HPAD employs a lightweight Pair-Agent debate framework with an adaptive stopping criterion to reduce token consumption. 
As shown in Table~\ref{tab:ablation_study_calibrated}, when escalating directly to voting after HCV, the setting  `w/o HPAD' decreases the accuracy from 86.30\% to 83.58\% on MMLU. 
In contrast, on the GPQA dataset, performance shows a slight increase from 54.04\% to 54.55\%; however, token consumption suddenly rises from 3,724 to 6,238. 
This increase is attributed to the complexity of the GPQA, where most queries cannot be effectively resolved by the debate mechanism alone and require voting with more agents. 
In conclusion, the pair-agent debate framework is an efficient and lightweight solution that minimizes token consumption while effectively completing the debate process.

\begin{table}[!t]
\centering
\fontsize{9pt}{9.5pt}\selectfont
\caption{Performance distribution across different stages of HCP-MAD on  MMLU. Rate (\%) denotes the fraction of queries finalized at HCV/HPAD, or escalated to ECV for final voting.}
\label{tab:stage_analysis}
\begin{tabular}{l|ccc}
\toprule
\textbf{Stages} & \textbf{Rate (\%)} & \textbf{Acc. (\%)} & \textbf{Tokens} \\
\midrule
HCV & 80.12 & 92.16 & 980 \\
HPAD & 14.31 & 64.29 & 5,695 \\
ECV & 5.57 & 62.20 & 6,736 \\
\midrule
\textbf{Overall} & \textbf{100} & \textbf{86.30} & \textbf{1977} \\
\bottomrule
\end{tabular}
\vspace{-1.3em}
\end{table}

%\paragraph{Effect of Progressive Reasoning}
\noindent\textbf{Effect of Progressive Reasoning.}
HCP-MAD applies a progressive reasoning mechanism to resolve simple problems quickly while allowing extensive reasoning on complex issues. 
As shown in Table~\ref{tab:stage_analysis}, 80.12\% of queries are successfully resolved at the initial HCV stage, achieving an accuracy of 92.16\% with 980 tokens. 
This indicates that consensus effectively addresses most queries. 
Additionally, 14.31\% of unresolved tasks are addressed at the HPAD stage, where the accuracy drops to 64.29\% and requires more tokens. 
This suggests that these tasks are more complex and that debate helps in resolving conflicts. 
Finally, the remaining 5.57\% of queries are tackled in the ECV stage, which results in the lowest accuracy of 62.20\% and the highest token consumption of 6,736. 
Overall, HCP-MAD effectively apply low-cost methods to the majority of simple tasks while reserving more resource-intensive voting for the few difficult ones.

\noindent\textbf{Effect of Early Stopping in HCV.} 
Heterogeneous Consensus Verification (HCV) utilizes an early stopping mechanism to prevent unnecessary reasoning processes, thereby reducing token consumption. 
As illustrated in Table~\ref{tab:ablation_study_calibrated}, disabling the early stopping mechanism, referred to as `w/o Early Stopping', not only increases token costs from 1,977 to 3,951 on the MMLU benchmark but also negatively impacts performance. 
This demonstrates that the consensus achieved by two heterogeneous agents in HCV serves as a reliable and computationally efficient criterion for stopping, allowing for reduced token consumption without sacrificing accuracy.

\noindent\textbf{Effect of Adaptive Stopping Criterion (ASC).} 
ASC is designed to dynamically terminate invalid debates, thereby reducing token consumption. 
As shown in Table~\ref{tab:ablation_study_calibrated}, ASC effectively decreases token usage from 2,753 to 1,997 on MMLU, while also increasing accuracy from 86.02\% to 86.30\%. 
This analysis shows that ASC optimizes the debate process by minimizing unnecessary computation and enhancing both efficiency and performance.

\noindent\textbf{Analysis of Escalated Collective Voting (ECV).} 
ECV is proposed to recruit additional agents to introduce broader collective intelligence for the complex tasks.
On the complex GPQA dataset, it improves the performance from 51.52\% to 54.04\%, while the token consumption has only slightly increased from 3,412 to 3,724.
Several critical aspects of ECV are analyzed below:
%\begin{itemize}

\textbf{1) Vote \emph{vs.} Debate.} 
For the final unresolved tasks, we employ a voting mechanism instead of a debate mechanism. 
As shown in Table~\ref{tab:ablation_study_calibrated}, substituting five-agent voting with a five-agent debate results in poorer performance, achieving only 52.53\% accuracy, along with a high token consumption of 4,698 on GPQA. 
This decline in performance can be attributed to answer exchange and persistent deadlock in HPAD, which reduce the effectiveness of debate in the final stage. 
\begin{table}
    \centering
\fontsize{9pt}{9.5pt}\selectfont
    
    \setlength{\tabcolsep}{4pt} % 稍微缩小列间距以适应宽度
        \caption{Effect of debate rounds and voting agents.}
    \begin{tabular}{c| c | c c | c c}
        \toprule
        \multirow{2}{*}{\textbf{Rounds}} & \multirow{2}{*}{\textbf{Agents}} & \multicolumn{2}{c|}{\textbf{MMLU}} & \multicolumn{2}{c}{\textbf{GPQA}} \\
          & & \textbf{Acc.} & \textbf{Tokens} & \textbf{Acc.} & \textbf{Tokens} \\
        \midrule
        2 & 5 & 86.02 & 1,730 & 55.05 & 4,131 \\
        4 & 3 & 86.16 & \textbf{1,574} & 53.03 & \textbf{3,283} \\
        4 & 5 & {86.30} & 1,709 & 54.04 & 3,724 \\
        4 & 7 & \textbf{86.57} & 2,073 & \textbf{56.57} & 4,767 \\
        6 & 5 & {86.30} & 1,894 & 50.00 & 3,849 \\
        \bottomrule
    \end{tabular}

    \label{tab:scaling_study}
\vspace{-1.0em}
\end{table}
\textbf{2) Weighted Voting \emph{vs.} Majority Voting.}
%As shown in Table~\ref{tab:ablation_study_calibrated}, 
`w/ Simple majority vote' means using only a simple majority voting from independent agents, which degrades the performance to  52.53\% on GPQA because they do not capture the important debate history context. 
Meanwhile, letting agents as debate reviewers vote `w/ Reviewer majority vote' achieves 53.03\% accuracy with a high cost of 4651 tokens, 
%as they may get stuck on historical bias. 
In contrast, our weighted group voting works best by leveraging the strengths of both groups, obtaining a superior performance of 54.04\% with a less token consumption of 3724 on the GPQA dataset. 
\textbf{3)Effect of Weighted Bonus.}
Removing the weighted bonus results in a worse performance on MMLU/GPQA datasets, \emph{e.g.,} 86.02\%/53.03\% \emph{vs} 86.30\%/54.04\%, showing its effectiveness in rewarding agents that provide more valuable insights.

% \begin{table}[t]
% \centering
% \small
% \setlength{\tabcolsep}{2.5pt}
% \renewcommand{\arraystretch}{0.92}

% \begin{tabular}{@{}ccrrrr@{}}
% \toprule
% \textbf{Debate} & \textbf{Voting}
% & \multicolumn{2}{c}{\textbf{MMLU}}
% & \multicolumn{2}{c}{\textbf{GPQA}} \\
% \cmidrule(lr){3-4}\cmidrule(lr){5-6}
% \textbf{Rounds} ($R$) & \textbf{Agents} ($N$)
% & \textbf{Acc.} & \textbf{Tokens}
% & \textbf{Acc.} & \textbf{Tokens} \\
% \midrule
% 2 & 5 & 86.02 & 1,730 & 55.05 & 4,131 \\
% 4 & 3 & 86.16 & \textbf{1,574} & 53.03 & \textbf{3,283} \\
% 4 & 5 & 86.30 & 1,977 & 54.04 & 3,724 \\
% 4 & 7 & \textbf{86.57} & 2,073 & \textbf{56.57} & 4,767 \\
% 6 & 5 & 86.30 & 1,894 & 50.00 & 3,849 \\
% \bottomrule
% \end{tabular}

% \caption{Effect of debate rounds and voting agents.}
% \label{tab:scaling_study}
% \vspace{-1.0em}
% \end{table}

\noindent\textbf{Scaling Study (Debate Rounds \& Voting Agent)}
Table~\ref{tab:scaling_study} reports the scaling results for voting agents and debate rounds.%(Table~\ref{tab:scaling_study})..%, and summarize the results in Table~\ref{tab:scaling_study}.
We can observe that setting 4 rounds obtains the better performance on both benchmarks.
The reason is that the existence of answer exchange and persistent deadlock makes a more debate round unnecessary. 
For the number of agents for voting in ECV, it can be observed that using five agents obtains the best performance on MMLU, while seven agents are suitable for GPQA because GPQA is more challenging than MMLU.

\section{Conclusion}
In this work, we propose a novel Heterogeneous Consensus-Progressive Reasoning for Efficient Multi-Agent Debate (HCP-MAD).
%Heterogeneous Consensus-Progressive Reasoning for Efficient Multi-Agent Debate that dynamically routes tasks by complexity. 
% an efficient multi-agent debate framework with heterogeneous consensus-driven progressive reasoning that routes tasks by difficulty.
HCP-MAD first checks rapid consensus for early stopping, then conducts a heterogeneous pair-agent debate with an adaptive stopping criterion, and escalates only complex tasks to collective voting with independent and reviewer agents. 
Experiments across benchmarks show higher accuracy with lower token costs. 
% Heterogeneous debates can resolve many queries, but complex problems may need more diverse agents. 
% Thus, finding ways to optimize topological dynamics to increase the number of debating agents without significantly raising token consumption will be our future work.
\newpage
\section*{Limitations}

Despite the effectiveness of HCP-MAD, our work has three key limitations:
(1) The evaluation is mainly conducted on reasoning and question-answering benchmarks; more complex scenarios can be further explored to verify its generalization.
(2) HCP-MAD relies on a heterogeneous agent pool. Although heterogeneity improves collaboration efficiency, how to select suitable agents for different tasks remains an important problem.
(3) The stopping and escalation rules  can be further optimized. Future work may explore more adaptive uncertainty-aware policies to better decide when to stop debate or escalate to collective voting.

\section*{Ethical Considerations}

HCP-MAD is a general framework for improving the efficiency of multi-agent reasoning, and the method itself does not introduce additional task-specific ethical risks. 
Since it relies on LLM-generated responses, it may still inherit biases or hallucinations from the underlying models. 
For high-stakes applications, human oversight is still necessary.
\bibliography{main}
\appendix % 开启附录模式
%\vspace{0.5\baselineskip}   
\section{HCP-MAD Algorithm}
We summarize the pipeline of HCP-MAD as the following algorithm.

\begin{algorithm}[]
    \caption{HCP-MAD Framework}
    \label{alg:HCP-MAD-compact}
    \small % 使用小号字体进一步节省空间
    \begin{algorithmic}[1]
        \STATE \textbf{Input:} Query $q$, Agents $\{a_1, \dots, a_N\}$, Max steps $T$, Model set $M$, System prompt $P_i$
        %\STATE \textbf{Note:} $\mathcal{J}(s)$ extracts answer $y_{i,t}$ from reasoning state $\boldsymbol{r}_{i,t}$
         \STATE \textbf{Output:} Final answer $\hat{y}$
        \STATE \textbf{Note:} The answer $\hat{y}_{i,t}$ is extracted from the reasoning state $\boldsymbol{r}_{i,t}$.
        %via extraction function $\mathcal{J}(\cdot)$.
        \STATE \textbf{Stage I: Initial Check ($t=0$)}
        %\STATE for $a_i \in \{a_1, a_2\}$, generate 
        \STATE $\boldsymbol{r}_{i,0} \gets \Psi_i(q, P_i, \emptyset)$ for each $a_i, i \in \{1, 2\}$
        %\STATE $\boldsymbol{r}_{i,0} \gets \Psi_i(q, P_i, \emptyset), i \in \{1, 2\}$; 
         \STATE \textbf{if} $\hat{y}_{1,0} = \hat{y}_{2,0}$ \textbf{then return} $\hat{y}_{1,0}$
        
        \STATE \textbf{Stage II: Heterogeneous Debate ($1 \le t < T$)}
        \FOR{$t = 1$ \TO $T-1$}
            %\STATE $\boldsymbol{r}_{i,t} \gets \Psi_i(q, P_i, \{\boldsymbol{r}_{1:2, <t}\}), i \in \{1, 2\}$
            \STATE $\mathcal{H}_{i,t} = \{\boldsymbol{r}_{i,t-1}, \boldsymbol{r}_{j,t-1}\}$ 
            \STATE $\boldsymbol{r}_{i,t} = \Psi_i(q, P_i, \mathcal{H}_{i,t})$ for each $a_i,
            i \in \{1, 2\}$          
            \STATE \textbf{if} $\hat{y}_{1,t} = \hat{y}_{2,t}$ \textbf{then return} $\hat{y}_{1,t}$ %\COMMENT{Early exit on consensus}
            \STATE \textbf{if} \text{ answer exchange} \textbf{or} \text{deadlock} \textbf{then break} %\COMMENT{Escalate next stage}
        \ENDFOR
        
        \STATE \textbf{Stage III: Escalated Voting ($t=T$)}
        %\STATE $\mathcal{C}_{deb} \gets \text{Summarize}(\{\boldsymbol{r}_{1:2, 0:T-1}\})$
        \STATE $\mathcal{C}_{deb} \gets Summarize(\boldsymbol{r}_{1,T-1},\boldsymbol{r}_{2,T-1})$
        \FORALL{$a_i \in \mathcal{A}_{esc} = \{a_3, \dots, a_N\} = \mathcal{A}_{ind} \cup \mathcal{A}_{rev}$}
            \STATE $H \gets \mathcal{C}_{deb}$ \textbf{if} $a_i \in \mathcal{A}_{rev}$ \textbf{else} $\emptyset$
            \STATE $\boldsymbol{r}_{i,T} \gets \Psi_i(q, P_{i}, H)$
        \ENDFOR
        \STATE \textbf{return} $\hat{y} = \text{WeightedMajorityVote}(\{\hat{y}_{i,T}\}_{i=3}^N, \{w_i\})$
    \end{algorithmic}
\end{algorithm}

\section{Experimental Details}
This section provides a comprehensive overview of our experimental setup, including dataset sizes, model configurations, and specific implementation parameters to ensure reproducibility. All experiments are conducted via API calls using the base models specified in the main text. \textbf{The source code will be released upon acceptance.}%The code is available at \url{https://anonymous.4open.science/r/002-2524/}.

\subsection{Dataset Description}
%\section{Dataset Details}
\label{sec:appendix_datasets}

Following prior work, we use the same dataset settings and evaluate our method on six benchmarks covering mathematical reasoning and commonsense knowledge. Table~\ref{tab:dataset_stats} summarizes the statistics and evaluation metrics for each dataset.

\begin{table}[h]
    \centering
    \small % 稍微缩小字体以适应栏宽
    \setlength{\tabcolsep}{4pt}
    \begin{tabular}{llc}
    \toprule
    \textbf{Dataset} & \textbf{Domain} & \textbf{Test Size} \\
    \midrule
    \multicolumn{3}{l}{\textit{Mathematical Reasoning}} \\
    GSM8K & Grade School Math & 1,319  \\
    MATH-500 & Competition Math & 500 \\
    AQuA & Algebra Problems & 254  \\
    \midrule
    \multicolumn{3}{l}{\textit{Commonsense \& General Knowledge}} \\
    MMLU & Academic Subjects & 1,474  \\
    CommonsenseQA & General Sense & 1,221  \\
    GPQA & Graduate Science & 198  \\
    \bottomrule
    \end{tabular}
    \caption{\textbf{Overview of Datasets.} We report the number of samples used for evaluation.}
    \label{tab:dataset_stats}
\end{table}

\subsection{Implementation Settings}
\label{sec:appendix_imp}
All experiments are conducted via API with a maximum generation limit of 2,048 tokens; for inference, we employ sampling ($T=0.7$, Top-$p=1.0$) for all diversity-based methods (SC, and MAD) while using greedy decoding ($T=0$) for standard CoT. To ensure fair resource allocation, baseline configurations are set as follows: Self-Consistency (SC) uses $k=3$ sampled paths, Self-Refine (SR) is limited to 2 iteration, standard MAD employs $N=3$ agents with $T=2$ fixed rounds, and MARS adopts a dual-author setup with a dedicated meta-reviewer. For our proposed HCP-MAD, we utilize an initial heterogeneous pair ($N_{init}=2$) and an escalation recruitment of $N_{esc}=5$ agents ($N_{ind}=2$ independent, $N_{rev}=3$ reviewers) across a maximum of $T=4$ debate rounds. Early stopping is triggered by consensus ($\Phi_{init}=1$), answer exchange ($\eta_e=2$), or deadlock ($\eta_d=2$ consecutive rounds with unchanged answers), while the Escalated Collective Voting (ECV) mechanism applies a base weight $w_{\text{base}} = 1.0$ and an independent consensus bonus, \[
\beta = \frac{N_{\text{rev}} - N_{\text{ind}}}{N_{\text{rev}}}
      = \frac{3 - 2}{3}
      = \frac{1}{3}.
\].

\section{Response Transitions Analysis} 
Table~\ref{tab:flip_analysis} summarizes the proportions of response transitions during the reasoning process. 
Overall, HCP-MAD exhibits a more favorable correction profile, as it more frequently corrects incorrect answers while rarely changing correct answers into incorrect ones. 
For example, HCP-MAD achieves a correction rate of 6.38\% for incorrect to correct transitions ( \textcolor{red}\xmark{} $\to$ \textcolor{green!60!black} \cmark ) compared to only 1.13\% for correct to incorrect transitions ( \textcolor{green!60!black} \cmark{} $\to$ \textcolor{red}\xmark ). 
Additionally, HCP-MAD maintains the accuracy of initial predictions with high stability, achieving a rate of 79.92\% for correct to correct transitions ( \textcolor{green!60!black} \cmark{} $\to$ \textcolor{green!60!black} \cmark ). 
In contrast, MAD shows significantly higher rates of harmful flips, such as 14.85\% for correct to incorrect transitions.%, indicating more instability in response shifts. 

\begin{table}
\centering
\fontsize{9pt}{9.5pt}\selectfont
\begin{tabular}{c|cccc}
\toprule
%待填写数据
 \textbf{Shift} & \textbf{MAD} & \textbf{SC} & \textbf{HCP-MAD} \\
\midrule
\textcolor{red}\xmark $\to$  \textcolor{red}\xmark & 12.23 & 17.90 & 12.57 \\
\textcolor{green!60!black} \cmark $\to$ \textcolor{red}\xmark &  14.85 &  1.75 &  1.13 \\
\textcolor{green!60!black}\cmark $\to$ \textcolor{green!60!black}\cmark & 66.81 & 76.42 & 79.92 \\
\textcolor{red}\xmark $\to$ \textcolor{green!60!black} \cmark&  6.11 &  3.93 &  6.38 \\
\bottomrule
\end{tabular}
    \caption{Breakdown of response transitions on the MMLU benchmark before and after applying multi-agent debate.% We use \cmark~and \xmark~to denote correct and incorrect answers, respectively. 
    }
\label{tab:flip_analysis}
\vspace{-1.0em}
\end{table}

\section{Additional Generalization Results}
\label{sec:appendix_generalization}

To assess the robustness and universality of our proposed framework, we extend our evaluation beyond the primary Llama-3.1 and Qwen2.5 pair. We introduce two additional heterogeneous agent pairs:
\begin{enumerate}
    \item \textbf{GPT-4o-mini + Mistral/Mixtral-8x22b-instruct}:
    Combining a cost-effective proprietary model with a strong Mixture-of-Experts (MoE) open model.
    \item \textbf{GPT-4.1-mini + Gemini-2.0-Flash}: Pairing two high-efficiency proprietary models from different providers (OpenAI and Google).
\end{enumerate}

Detailed performance comparisons on all six benchmarks are presented in Table~\ref{tab:performance_comparison1} and Table~\ref{tab1:performance_comparison2}. The results demonstrate that HCP-MAD consistently outperforms single-agent baselines and standard consensus methods across these diverse architectures, validating that the effectiveness of our consensus-based stopping mechanism is not dependent on specific model sets. Furthermore, our results highlight a superior acc-cost trade-off.

%\clearpage
\section{Case Studies}
\label{sec:appendix_cases}
We present representative examples from GPQA and GSM8K to demonstrate the adaptive workflow of HCP-MAD across varying levels of problem complexity and different failure modes.

% --- Case 1: Easy (Consensus) ---
\begin{tcolorbox}[
    enhanced, breakable, width=\linewidth,
    colback=green!5, colframe=green!40!black, % Light green for "Easy/Success"
    title=\textbf{Case 1: Early Consensus (Easy Task) - GPQA ID 2},
    fonttitle=\bfseries\small
]
\textbf{Question:}
A spin-half particle is in a linear superposition
$0.5\,\ket{\uparrow}+\frac{\sqrt{3}}{2}\,\ket{\downarrow}$
of its spin-up and spin-down states.
If $\ket{\uparrow}$ and $\ket{\downarrow}$ are the eigenstates of $\sigma_{z}$,
then what is the expectation value (up to one decimal place) of the operator
$10\sigma_{z}+5\sigma_{x}$?
Here, symbols have their usual meanings:\\
A) 1.65, B) 0.85, C) -0.7, D) -1.4.\\
\textbf{Ground Truth:} C

\vspace{4pt}
\textbf{Stage 1 (HCV) - Initial Check:}
\begin{itemize}[leftmargin=1.5em, nosep]
    \item \textbf{Agent 1 (Round 0):} ``Based on ..., the answer is C.'' $\to$ \textbf{C}
    \item \textbf{Agent 2 (Round 0):} ``The calculation shows ..., therefore C.'' $\to$ \textbf{C}
\end{itemize}

\vspace{4pt}
\textbf{Outcome:} Consensus ($\hat{y}_1 = \hat{y}_2$) reached at $t=0$. \textbf{Early Stop triggered.} \\
\textbf{Result:} \textcolor{green!60!black}{\textbf{Correct (C)}}. \\
\textit{Analysis: For straightforward queries, the heterogeneous pair efficiently verifies the consensus without incurring the cost of debate.}
\end{tcolorbox}

\vspace{10pt}

% --- Case 2: Medium (Debate) ---
\begin{tcolorbox}[
    enhanced, breakable, width=\linewidth,
    colback=blue!5, colframe=blue!40!black, % Light blue for "Medium/Debate"
    title=\textbf{Case 2: Resolution via Debate (Medium Task) - GPQA ID 6},
    fonttitle=\bfseries\small
]
\textbf{Question:}
The universe is filled with the Cosmic Microwave Background (CMB). Consider
$\gamma\gamma\rightarrow e^{+}e^{-}$ between a high-energy $\gamma$-ray and a CMB photon.
From what $\gamma$-ray energy would lifetimes be limited by this process?
Given $\langle \epsilon_{\rm CMB}\rangle = 10^{-3}\,\mathrm{eV}$, \dots\ 
\\A) $1.8\times10^{5}$ GeV,\\\
B) $2.6\times10^{5}$ GeV,\\
C) $3.9\times10^{5}$ GeV,\\
D) $9.5\times10^{4}$ GeV.\\
\textbf{Ground Truth:} B

\vspace{4pt}
\textbf{Stage 1 (HCV) - Initial Check:}
\begin{itemize}[leftmargin=1.5em, nosep]
    \item \textbf{Round 0:} Agent 1 $\to$ \textbf{A}; Agent 2 $\to$ \textbf{C}.
\end{itemize}
\textit{Status: Disagreement. Proceed to Debate.}

\vspace{4pt}
\textbf{Stage 2 (HPAD) - Heterogeneous Debate:}
\begin{itemize}[leftmargin=1.5em, nosep]
    \item \textbf{Round 1:} 
    \begin{itemize}[nosep]
        \item Agent 1: Shifts to \textbf{B}.
        \item Agent 2: Shifts to \textbf{A}.
    \end{itemize}
    \item \textbf{Round 2:}
    \begin{itemize}[nosep]
        \item Agent 1: ``Re-evaluating supports B...'' $\to$ \textbf{B}
        \item Agent 2: ``The argument for B is compelling...'' $\to$ \textbf{B}
    \end{itemize}
\end{itemize}

\vspace{4pt}
\textbf{Outcome:} Consensus reached at $t=2$. \\
\textbf{Result:} \textcolor{green!60!black}{\textbf{Correct (B)}}. \\
\textit{Analysis: The debate allowed agents to correct initial hallucinations (A and C) and converge on the correct solution (B) through iterative critique.}
\end{tcolorbox}

\vspace{10pt}

\begin{table*}[]
\centering
\small % 缩小字号以适应页面宽度
\setlength{\tabcolsep}{6pt}
\begin{tabular}{c|cccccc|cc}
\toprule
Methods & MMLU & CommonQA & GPQA & MATH500 & GSM8K &  AQuA & \textbf{Avg Acc.(\%)} & \textbf{Avg Tokens} \\
\midrule
\multicolumn{9}{>{\columncolor{gray!15}}c}{GPT-4o-mini} \\
\midrule
CoT & 80.12 & 81.16 & 41.92 & 72.60 & 92.87 & 80.31 & 74.83 &518 \\
SR & 75.98 & 77.56 & 38.38 & 71.80 & 91.13 & 78.35 & 72.20 & 1,191 \\
SC & 81.07 & 81.49 & 43.94 & 75.80 & 94.16 & 82.68 &  \underline{79.86} &1,539 \\
\midrule
%MAD & 77.95 & 78.30 & 49.49 & 82.20 & 91.89 & 86.61 & 77.74 & 7,413 \\
%\midrule
\multicolumn{9}{>{\columncolor{gray!15}}c}{Mistral/Mixtral-8x22b-instruct} \\
\midrule
CoT & 80.19 & 80.84 & 45.96 & 73.40 & 94.39 & 81.89 & 76.11 &567 \\
SR & 78.22 & 73.96 & 43.94 & 75.00 & 94.69 & 83.07 & 74.81 & 1,214 \\
SC & 80.12 & 80.67 &{46.97} & 76.00 & 93.25 & 83.86 & 76.81 & 1,876 \\
\midrule
MAD & \underline{81.82} & \underline{81.98} & 45.96 & \underline{77.40} & \underline{95.38} & \underline{84.65} & 77.87 & 6,857 \\
\midrule
\multicolumn{9}{>{\columncolor{gray!15}}c}{GPT-4o-mini + Mistral/Mixtral-8x22b-instruct} \\
\midrule
MARS & 78.56 & 81.82 & 45.45 & 73.80 & 90.67 & 78.74 & 74.84 & 3,754 \\
Heter-MAD & 75.58 & 73.79 &  \underline{47.98}& 74.80 & 92.65 & 65.35 & 73.35 &3,427\\
DOWN & 75.64
& 67.16
& 38.38
& 61.80
& 93.25
& 83.07&
75.73&2,579\\
HCP-MAD & \textbf{84.06} & \textbf{84.03} & \textbf{55.05} & \textbf{83.60} & \textbf{96.06} & \textbf{87.40} & \textbf{81.87} & 2,540 \\
\bottomrule
\end{tabular}
\caption{Performance and cost-efficiency comparison on the {GPT-4o-mini + Mixtral-8x22B-Instruct} heterogeneous pair. Best scores are highlighted in {bold}, and second-best are \underline{underlined}. Scores for all benchmarks and Avg Acc. are reported as accuracy percentages (\%). Avg Tokens denotes the average number of tokens consumed per query.}
\label{tab:performance_comparison1}
\vspace{-1.0em}
\end{table*}

\begin{table*}[]
\centering
\small % 缩小字号以适应页面宽度
\setlength{\tabcolsep}{6pt}
\begin{tabular}{c|cccccc|cc}
\toprule
Methods & MMLU & CommonQA & GPQA & MATH500 & GSM8K &  AQuA & \textbf{Avg Acc.(\%)} & \textbf{Avg Tokens} \\
\midrule
\multicolumn{9}{>{\columncolor{gray!15}}c}{GPT-4.1-mini } \\

\midrule
CoT & 87.31 & 82.88 & 64.65 & 87.80 & 96.06 & 88.58 & 84.81 & 604 \\
SR  & 87.25 & 80.43 & 67.68 & 87.60 & 94.47 & 89.37 & 84.47 & 1,547 \\
SC  & 86.97 &\textbf{ 84.11} & 65.66 & \underline{89.80} & \textbf{96.06} & 88.98 & \underline{85.26} & 1,824 \\
\midrule
% MAD & 77.95 & 78.30 & 49.49 & 82.20 & 91.89 & 86.61 & 77.74 & 7,413 \\
% \midrule
\multicolumn{9}{>{\columncolor{gray!15}}c}{Gemini-2.0-Flash} \\
\midrule
CoT & 86.70 & 83.13 & 61.62 & 86.40 & 95.60 & 86.61 & 83.34 & 564 \\
SR  & 82.90 & 77.72 & 62.12 & 88.60 & 94.84 & 88.58 & 82.46 & 1,228 \\
SC  & 87.25 & 82.47 & 65.15 & 88.20 & \underline{95.91} & \textbf{89.76} & 84.79 & 1,745 \\
\midrule
MAD & \underline{88.13} & 83.87 & 65.15 & 88.60 & 94.77 & 88.98 & 84.92 & 6,621 \\
\midrule
\multicolumn{9}{>{\columncolor{gray!15}}c}{GPT-4.1-mini + Gemini-2.0-Flash} \\
\midrule
MARS  & 86.57 & 81.41 & 66.67 & 88.40 & 94.01 & 86.22 & 83.88 & 3,567 \\
Heter-MAD & 85.48 & 82.80&\underline{68.69} & 71.20& 91.74 & 83.07 & 80.50 & 3,168 \\
DOWN & 83.11
& 77.15
& 62.12
& 70.00
& 95.53
& 88.98
& 83.08&3,219 \\
HCP-MAD & \textbf{88.26} & \underline{84.03} & \textbf{73.74} & \textbf{91.00} & {95.53} & \textbf{89.76} & \textbf{87.05} & {2,816}  \\
\bottomrule
\end{tabular}
%\caption{Performance comparison across multiple datasets among single-agent methods, multi-agent debate (MAD) systems, and our proposed heterogeneous approach. Best scores are highlighted in bold. Avg Token represents the average computational cost per query.}
% ChatEval
\caption{Generalization results on the {GPT-4.1-mini + Gemini-2.0-Flash} pair. Best scores are highlighted in \textbf{bold}, and second-best are \underline{underlined}. Scores for all benchmarks and Avg Acc. are reported as accuracy percentages (\%). Avg Tokens denotes the average number of tokens consumed per query.
%Avg Token represents the average computational cost per query.
}
\label{tab1:performance_comparison2}
%\end{table}
\vspace{-1.0em}
\end{table*}

% --- Case 3: Hard (Escalation) ---
\begin{tcolorbox}[
    enhanced, breakable, width=\linewidth,
    colback=red!5, colframe=red!40!black, % Light red for "Hard/Escalation"
    title=\textbf{Case 3: Escalated Arbitration (Hard Task) - GPQA ID 109},
    fonttitle=\bfseries\small
]
\textbf{Question:}
Which stars can be detected with \emph{both} ESPRESSO ($V<17$) and HIRES ($V<16$),
ignoring pointing limits? \dots\ 
Star parameters:
(1) RA $=15^\circ$, DEC $=-75^\circ$, $M_V=15.5$, $d=10$ pc;
(2) RA $=30^\circ$, DEC $=55^\circ$, $m_V=16.5$, $d=5$ pc;
(3) RA $=11$ h, DEC $=48^\circ$, $m_V=15.5$, $E(B\!-\!V)=0.6$, $d=15$ pc;
(4) RA $=85^\circ$, DEC $=-48^\circ$, $M_V=15.5$, $E(B\!-\!V)=0.4$, $d=10$ pc;
(5) RA $=10$ h, DEC $=60^\circ$, $M_V=16.5$, $d=5$ pc;
with $A_V=3.1\,E(B\!-\!V)$.
Options:\\
A) Star2 and Star3,\\
B) Star1 and Star4,\\
C) Star4 and Star5,\\ 
D) Star3 and Star5.\\
\textbf{Ground Truth:} D

\vspace{4pt}
\textbf{Stage 1 (HCV) - Initial Check:}
\begin{itemize}[leftmargin=1.5em, nosep]
    \item \textbf{Round 0:} Agent 1 $\to$ \textbf{B}; Agent 2 $\to$ \textbf{D}.
\end{itemize}

\vspace{4pt}
\textbf{Stage 2 (HPAD) - Heterogeneous Debate:}
\begin{itemize}[leftmargin=1.5em, nosep]
    \item \textbf{Round 1:} Agent 1 $\to$ \textbf{D}; Agent 2 $\to$ \textbf{B}. (Flip)
    \item \textbf{Round 2:} Agent 1 $\to$ \textbf{A}; Agent 2 $\to$ \textbf{D}. 
    \item \textbf{Round 3:} Agent 1 $\to$ \textbf{C}; Agent 2 $\to$ \textbf{B}.
\end{itemize}
\textit{Status: Max rounds reached ($t=T-1$) without consensus. \textbf{Trigger ECV.}}

\vspace{4pt}
\textbf{Stage 3 (ECV) - Escalated Collective Voting:}
\begin{itemize}[leftmargin=1.5em, nosep]
    \item \textbf{Independent Group ($N=2$):}
    \begin{itemize}[nosep]
        \item Ind\_1: ``Calculating magnitudes...'' $\to$ \textbf{D}
        \item Ind\_2: ``Star 2 is too faint...'' $\to$ \textbf{B}
    \end{itemize}
    \item \textbf{Reviewer Group ($N=3$):}
    \begin{itemize}[nosep]
        \item Rev\_1 $\to$ \textbf{D}; Rev\_2 $\to$ \textbf{D}; Rev\_3 $\to$ \textbf{D}
    \end{itemize}
\end{itemize}

\vspace{4pt}
\textbf{Outcome:} Weighted Majority Vote: \textbf{D} (4 votes) vs \textbf{B} (1 vote). \\
\textbf{Result:} \textcolor{green!60!black}{\textbf{Correct (D)}}. \\
\textit{Analysis: The initial pair of agents was stuck in a cycle of incorrect answers. By adding independent observers and contextual reviewers, the escalation stage helped identify the correct reasoning and solved the difficult question.}
\end{tcolorbox}

\vspace{10pt}

% --- Case 4: Answer Exchange (Loop) ---
\begin{tcolorbox}[
    enhanced, breakable, width=\linewidth,
    colback=yellow!5, colframe=orange!60!black, % Orange for Mechanism Warning
    title=\textbf{Case 4: Answer Exchange (Loop Detection) - GSM8K ID 40},
    fonttitle=\bfseries\small
]
\textbf{Question:}
Brandon's iPhone is four times as old as Ben's iPhone. Ben's iPhone is two times older than Suzy's iPhone. If Suzy\u2019s iPhone is 1 year old, how old is Brandon\u2019s iPhone?\\
\textbf{Ground Truth:} 8.0

\vspace{4pt}
\textbf{Stage 1 (HCV) - Initial Check:}
\begin{itemize}[leftmargin=1.5em, nosep]
    \item \textbf{Round 0:} Agent 1 $\to$ \textbf{12.0}; Agent 2 $\to$ \textbf{8.0}.
\end{itemize}

\vspace{4pt}
\textbf{Stage 2 (HPAD) - Heterogeneous Debate:}
\begin{itemize}[leftmargin=1.5em, nosep]
    \item \textbf{Round 1 (Exchange):} 
    \begin{itemize}[nosep]
        \item Agent 1: ``Agree with Agent 2.'' $\to$ \textbf{8.0}
        \item Agent 2: ``Agree with Agent 1.'' $\to$ \textbf{12.0}
    \end{itemize}
    \item \textbf{Round 2 (Exchange):} 
    \begin{itemize}[nosep]
        \item Agent 1: ``Agree with Agent 2.'' $\to$ \textbf{8.0}
        \item Agent 2: ``Agree with Agent 1.'' $\to$ \textbf{12.0}
    \end{itemize}
\end{itemize}
\textit{Status: Answers swapped ($A, B \to B, A$). $\mathcal{E}_t \ge \eta_e$ detected. \textbf{Forced Break.}}

\vspace{4pt}
\textbf{Stage 3 (ECV) - Escalated Collective Voting:}
\begin{itemize}[leftmargin=1.5em, nosep]
    \item \textbf{Escalation Pool ($N=5$):}
    \begin{itemize}[nosep]
        \item Votes: [8.0, 8.0, 8.0, 12.0, 8.0]
    \end{itemize}
\end{itemize}

\vspace{4pt}
\textbf{Outcome:} Consensus on \textbf{8.0} (4 votes) vs 12.0 (1 votes). \\
\textbf{Result:} \textcolor{green!60!black}{\textbf{Correct (8.0)}}. \\
\textit{Analysis: This mechanism stops repetitive answer swapping. Escalation introduces new viewpoints and further checks to break the cycle and verify the final answer.}
\end{tcolorbox}

\vspace{10pt}

% --- Case 5: Deadlock Resolution ---
\begin{tcolorbox}[
    enhanced, breakable, width=\linewidth,
    colback=red!5, colframe=red!40!black, % Red for Deadlock
    title=\textbf{Case 5: Deadlock Resolution (Stalemate) - GSM8K ID 242},
    fonttitle=\bfseries\small
]
\textbf{Question:}
Mike was a pen pal with 5 people.  He stopped being penpals with 2 of them.  They each send 2 letters a week that are 5 pages long.  He responds in kind.  He can write a page every 6 minutes.  How many hours does he spend writing a week?\\
\textbf{Ground Truth:} 3.0

\vspace{4pt}
\textbf{Stage 1 (HCV) - Initial Check:}
\begin{itemize}[leftmargin=1.5em, nosep]
    \item \textbf{Round 0:} Agent 1 $\to$ \textbf{6.0}; Agent 2 $\to$ \textbf{3.0}.
\end{itemize}

\vspace{4pt}
\textbf{Stage 2 (HPAD) - Heterogeneous Debate:}
\begin{itemize}[leftmargin=1.5em, nosep]
    \item \textbf{Round 1:} Agent 1 $\to$ \textbf{6.0}; Agent 2 $\to$ \textbf{3.0}. (No change)
    \item \textbf{Round 2:} Agent 1 $\to$ \textbf{6.0}; Agent 2 $\to$ \textbf{3.0}. (Stubborn)
\end{itemize}
\textit{Status: Deadlock counter $\mathcal{D}_t \ge \eta_d$. Debate is stuck. \textbf{Trigger Escalation.}}

\vspace{4pt}
\textbf{Stage 3 (ECV) - Escalated Collective Voting:}
\begin{itemize}[leftmargin=1.5em, nosep]
    \item \textbf{Independent Group ($N=2$):} Both derived \textbf{3.0}.
    \item \textbf{Reviewer Group ($N=3$):} 2 of 3 supported \textbf{3.0}.
    \item \textbf{Final Vote Distribution:} [6.0, 3.0, 3.0, 3.0, 3.0]
\end{itemize}

\vspace{4pt}
\textbf{Outcome:} Overwhelming consensus on \textbf{3.0} (4 votes). \\
\textbf{Result:} \textcolor{green!60!black}{\textbf{Correct (3.0)}}. \\
\textit{Analysis: When two agents are confidently wrong (or one is right but unpersuasive), standard debate can fail. ECV mitigates this by aggregating additional independent judgments, which increases the chance of surfacing and selecting the correct answer.}
\end{tcolorbox}

\section{Prompt Templates}
This section details the prompt templates employed across the three stages of the HCP-MAD framework. These templates are designed to manage agent interactions during the heterogeneous debate, and coordinate the final escalation voting process. %We provide specific instructions for initial answer generation, stability-focused debate rounds, and the dual-role escalation mechanism (comprising both unbiased observers and contextual reviewers) to ensure the transparency and reproducibility of our multi-agent coordination system.
% =======================================================
% Cross two-column Prompt display
% =======================================================
\begin{figure*}[t]
    \centering

    % --- Prompt 1: Initial Answer ---
    \begin{promptbox}{Prompt Stage 1: Heterogeneous Consensus Verification (Initial Answer)}
        \footnotesize
        \setlength{\parskip}{2pt}
        \setlength{\parindent}{0pt}

        \textbf{Input:} \texttt{\{{Task\_Instruction\}}} + \texttt{\{{User\_Query\}}}

        \vspace{3pt}\hrule\vspace{3pt}

        \textbf{Output Requirements:}
        \begin{itemize}[leftmargin=1.5em]
            \item \textbf{Mathematical Tasks:}
            ``You are a math assistant. Please help to solve the following math problem: [user\_query]
            Give your thoughts about the computation steps and the final numerical answer...
            Final answer must be a single numerical number.''

            \item \textbf{Multiple-Choice:}
            ``You are an assistant. Please help to solve the following problem: [user\_query]
            Give your thoughts... Answer: (X) where X is chosen from [A,B,C,D].''

            % \item \textbf{Code Generation (HumanEval, MBPP):}
            % ``You are an expert Python programmer. Complete the following Python function...
            % Only provide the function implementation.''

            % \item \textbf{MATH/MATH500:}
            % ``Give your step-by-step solution and put your final answer in \texttt{$\backslash$boxed\{...\}}.''
        \end{itemize}
    \end{promptbox}

    % --- Prompt 2: Debate (V3) ---
    \begin{promptbox}{Prompt Stage 2: Heterogeneous Pair-Agent Debate}
        \footnotesize
        \setlength{\parskip}{2pt}
        \setlength{\parindent}{0pt}

        \textbf{Round 0 (Independent Answer):} Uses the same format as Stage 1.

        \vspace{3pt}\hrule\vspace{3pt}

        \textbf{Subsequent Rounds:}

        Problem: \texttt{\{{User\_Query\}}}

        \vspace{2pt}

        \textbf{=== YOUR PREVIOUS POSITION ===} \\
        Your previous answer: \texttt{\{{My\_Previous\_Answer\}}} \textit{(stable for N rounds)}

        \vspace{2pt}

        \textbf{=== OTHER AGENT'S SOLUTION ===} \\
        Their answer: \texttt{\{{Other\_Agent\_Answer\}}} \\
        Their reasoning: \texttt{\{{Other\_Agent\_Reasoning\}}}

        \vspace{2pt}

        \textbf{=== STABILITY NOTE (if stable $\geq$ 2 rounds) ===} \\
        You have maintained the same answer for N consecutive rounds.
        This suggests high confidence. Be extra cautious about changing.

        \vspace{2pt}

        \textbf{=== DECISION RULE (Crucial for Stability) ===}
        \begin{itemize}[leftmargin=1.5em]
            \item You may \textbf{ONLY} change your answer if you find a \textbf{SPECIFIC ERROR} in your own reasoning.
            \item Do \textbf{NOT} change just because the other agent disagrees or sounds confident.
            \item If unsure, \textbf{KEEP} your original answer.
        \end{itemize}

        \vspace{2pt}

        \textbf{Required Output Format:} \\
        \texttt{Thoughts: [your step-by-step verification]} \\
         \texttt{Decision: [KEEP / CHANGE] - [specific error found OR no error found]} \\
        \texttt{Answer: [final answer in required format]}
    \end{promptbox}
    %\caption{“Prompt for Stage 1–2”}
    \label{fig:all_prompts_a}
\end{figure*}

\begin{figure*}[t]
    \centering
    % --- Prompt 3a: Escalation Independent Observers ---
    \begin{promptbox}{Prompt Stage 3a: Escalated Collective Voting (Independent Observers)}
        \footnotesize
        \setlength{\parskip}{2pt}
        \setlength{\parindent}{0pt}

        \textbf{Role:} Independent Observer (reasons from scratch, no debate history)

        \vspace{3pt}\hrule\vspace{3pt}

        \textbf{Input:} \texttt{\{{Task\_Instruction\}}} + \texttt{\{{User\_Query\}}}

        \vspace{2pt}

        \textit{Note: Independent Observers use the same prompt format as Stage 1 (Initial Answer),
        reasoning independently without access to the debate history. This ensures unbiased evaluation
        from a fresh perspective.}

        \vspace{2pt}

        \textbf{Output Format:} Same as Stage 1 (task-specific format)
    \end{promptbox}

    % --- Prompt 3b: Escalation Contextual Reviewers ---
    \begin{promptbox}{Prompt Stage 3b: Escalated Collective Voting (Contextual Reviewers)}
        \footnotesize
        \setlength{\parskip}{2pt}
        \setlength{\parindent}{0pt}

        \textbf{Role:} Evaluator reviewing a debate between two agents who could not reach consensus.

        \vspace{3pt}\hrule\vspace{3pt}

        \textbf{=== PROBLEM ===} \\
        \texttt{\{{User\_Query\}}}

        \vspace{2pt}

        \textbf{=== DEBATE SUMMARY ===} \\
        \texttt{\{{Debate\_Summary\_Text\}}} \\
        \textit{(Includes: Agent answers, logical stalemate counts, Last round arguments)}

        \vspace{2pt}

        \textbf{=== EVALUATION GUIDELINES ===}
        \begin{enumerate}[leftmargin=1.5em]
            \item A ``Persistent Deadlock'' agent (never changed answer) may be more confident in their reasoning.
            \item A ``Answerexchange'' agent may have been incorrectly persuaded OR correctly recognized an error.
            \item Focus on the \textbf{REASONING quality}, not just which agent seems more confident.
            \item Verify calculations/facts independently if possible.
        \end{enumerate}

        \vspace{2pt}

        \textbf{=== YOUR TASK ===} \\
        Based on the debate summary, determine the correct answer.

        \vspace{2pt}

        \textbf{Output Format:} \\
        \texttt{Thoughts: [your independent analysis]} \\
        \texttt{Answer: [final answer in required format]}
    \end{promptbox}

\caption{\textbf{Core Prompt Templates used in HCP-MAD.} \textbf{Stage 1} focuses on generating initial independent reasoning paths from heterogeneous agents. \textbf{Stage 2} facilitates iterative debate rounds where agents exchange and refine their answers. \textbf{Stage 3} implements an escalation mechanism involving two distinct groups: \textit{Independent Observers} (3a) who solve the problem from scratch to provide unbiased perspectives, and \textit{Contextual Reviewers} (3b) who analyze the previous debate history to identify and resolve logical conflicts.}

    \label{fig:all_prompts}
\end{figure*}

\end{document}